\documentclass[10pt,journal,amsmath,amsfonts,amssymb]{IEEEtran}

\usepackage{graphicx}
\usepackage{psfrag}

\long\def\symbolfootnote[#1]#2{\begingroup%
\def\thefootnote{\fnsymbol{footnote}}\footnote[#1]{#2}\endgroup}

\begin{document}

\title{Achieving the Gaussian Rate-Distortion Function by
Prediction}

\author{
\authorblockN{Ram Zamir, Yuval Kochman and Uri Erez}
\authorblockA{Dept. Electrical Engineering-Systems, Tel Aviv University \\
} }

\maketitle

\footnotetext{The work of the first two authors was partially
supported by the Israel Science Foundation, grant ISF 1259/07}

%
%
%
%
%

\begin{abstract}
The ``water-filling'' solution for the quadratic rate-distortion
function of a stationary Gaussian source is given in terms of its
power spectrum. This formula naturally lends itself to a frequency
domain ``test-channel'' realization. We provide an alternative
time-domain realization for the rate-distortion function, based on
linear prediction. The predictive test-channel has some interesting implications,
including the optimality at all distortion levels of pre/post
filtered vector-quantized differential pulse code modulation
(DPCM), and a duality relationship with decision-feedback
equalization (DFE) for inter-symbol interference (ISI) channels.
\end{abstract}



\newcommand{\ebj}{e^{j 2 \pi f}}
\newcommand{\uzer}{\underline{0}}
\newcommand{\uV}{\underline{V}}
\newcommand{\uA}{\underline{A}}
\newcommand{\uD}{\underline{D}}
\newcommand{\uv}{\underline{v}}
\newcommand{\uT}{\underline{T}}
\newcommand{\ut}{\underline{t}}
\newcommand{\ur}{\underline{r}}
\newcommand{\uR}{\underline{R}}
\newcommand{\uc}{\underline{c}}
\newcommand{\uC}{\underline{C}}
\newcommand{\ul}{\underline{l}}
\newcommand{\uL}{\underline{L}}
\newcommand{\uh}{\underline{h}}
\newcommand{\uH}{\underline{H}}
\newcommand{\ue}{\underline{e}}
\newcommand{\uE}{\underline{E}}
\newcommand{\uG}{\underline{G}}
\newcommand{\ug}{\underline{g}}
\newcommand{\uz}{\underline{z}}
\newcommand{\uZ}{\underline{Z}}
\newcommand{\uu}{\underline{u}}
\newcommand{\uU}{\underline{U}}
\newcommand{\uj}{\underline{j}}
\newcommand{\uJ}{\underline{J}}
\newcommand{\uX}{\underline{X}}
\newcommand{\ux}{\underline{x}}
\newcommand{\uY}{\underline{Y}}
\newcommand{\uy}{\underline{y}}
\newcommand{\uW}{\underline{W}}
\newcommand{\uw}{\underline{w}}
\newcommand{\uth}{\underline{\theta}}
\newcommand{\uTh}{\underline{\Theta}}
\newcommand{\uph}{\underline{\phi}}
\newcommand{\ual}{\underline{\alpha}}
\newcommand{\uxi}{\underline{\xi}}
\newcommand{\us}{\underline{s}}
\newcommand{\uS}{\underline{S}}
\newcommand{\un}{\underline{n}}
\newcommand{\uN}{\underline{N}}
\newcommand{\up}{\underline{p}}
\newcommand{\uq}{\underline{q}}
\newcommand{\uf}{\underline{f}}
\newcommand{\ua}{\underline{a}}
\newcommand{\ub}{\underline{b}}
\newcommand{\uDelta}{\underline{\Delta}}
\newcommand{\cA}{{\cal A}}
\newcommand{\cB}{{\cal B}}
\newcommand{\cC}{{\cal C}}
\newcommand{\cc}{{\cal c}}
\newcommand{\cD}{{\cal D}}
\newcommand{\cE}{{\cal E}}
\newcommand{\cI}{{\cal I}}
\newcommand{\cK}{{\cal K}}
\newcommand{\cL}{{\cal L}}
\newcommand{\cN}{{\cal N}}
\newcommand{\cP}{{\cal P}}
\newcommand{\cQ}{{\cal Q}}
\newcommand{\cR}{{\cal R}}
\newcommand{\cS}{{\cal S}}
\newcommand{\cs}{{\cal s}}
\newcommand{\cT}{{\cal T}}
\newcommand{\ct}{{\cal t}}
\newcommand{\cU}{{\cal U}}
\newcommand{\cV}{{\cal V}}
\newcommand{\cW}{{\cal W}}
\newcommand{\cX}{{\cal X}}
\newcommand{\cx}{{\cal x}}
\newcommand{\cY}{{\cal Y}}
\newcommand{\cy}{{\cal y}}
\newcommand{\cZ}{{\cal Z}}
\newcommand{\tE}{\tilde{E}}
\newcommand{\tZ}{\tilde{Z}}
\newcommand{\tz}{\tilde{z}}
\newcommand{\hU}{\hat{U}}
\newcommand{\hX}{\hat{X}}
\newcommand{\hY}{\hat{Y}}
\newcommand{\hZ}{\hat{Z}}
\newcommand{\huX}{\hat{\uX}}
\newcommand{\huY}{\hat{\uY}}
\newcommand{\huZ}{\hat{\uZ}}
\newcommand{\indp}{\underline{\; \| \;}}
\newcommand{\diag}{\mbox{diag}}
\newcommand{\sumk}{\sum_{k=1}^{K}}
\newcommand{\beq}[1]{\begin{equation}\label{#1}}
\newcommand{\eeq}{\end{equation}}
\newcommand{\req}[1]{(\ref{#1})}
\newcommand{\eqref}[1]{(\ref{#1})}
\newcommand{\figref}[1]{Figure \ref{#1}}
\newcommand{\secref}[1]{Section \ref{#1}}
\newcommand{\thref}[1]{Theorem \ref{#1}}
\newcommand{\beqn}[1]{\begin{eqnarray}\label{#1}}
\newcommand{\eeqn}{\end{eqnarray}}
\newcommand{\limn}{\lim_{n \rightarrow \infty}}
\newcommand{\limN}{\lim_{N \rightarrow \infty}}
\newcommand{\limr}{\lim_{r \rightarrow \infty}}
\newcommand{\limd}{\lim_{\delta \rightarrow \infty}}
\newcommand{\limM}{\lim_{M \rightarrow \infty}}
\newcommand{\limsupn}{\limsup_{n \rightarrow \infty}}
\newcommand{\imii}{\int_{-\infty}^{\infty}}
\newcommand{\imix}{\int_{-\infty}^x}
\newcommand{\ioi}{\int_o^\infty}
\newcommand{\vecN}{_0^{N-1}}

\newcommand{\bphi}{\mbox{\boldmath \begin{math}\phi\end{math}}}
\newcommand{\bth}{\mbox{\boldmath \begin{math}\theta\end{math}}}
\newcommand{\bhth}{\mbox{\boldmath \begin{math}\hat{\theta}\end{math}}}
\newcommand{\bg}{\mbox{\boldmath \begin{math}g\end{math}}}
\newcommand{\ba}{{\bf a}}
\newcommand{\bb}{{\bf b}}
\newcommand{\bc}{{\bf c}}
\newcommand{\bD}{{\bf D}}
\newcommand{\bbf}{{\bf f}}
\newcommand{\bn}{{\bf n}}
\newcommand{\bs}{{\bf s}}
\newcommand{\bt}{{\bf t}}
\newcommand{\bu}{{\bf u}}
\newcommand{\bx}{{\bf x}}
\newcommand{\by}{{\bf y}}
\newcommand{\bz}{{\bf z}}
\newcommand{\bC}{{\bf C}}
\newcommand{\bJ}{{\bf J}}
\newcommand{\bN}{{\bf N}}
\newcommand{\bQ}{{\bf Q}}
\newcommand{\bS}{{\bf S}}
\newcommand{\bT}{{\bf T}}
\newcommand{\bV}{{\bf V}}
\newcommand{\bX}{{\bf X}}
\newcommand{\bY}{{\bf Y}}
\newcommand{\bZ}{{\bf Z}}
\newcommand{\oI}{\overline{I}}
\newcommand{\oD}{\overline{D}}
\newcommand{\oh}{\overline{h}}
\newcommand{\oV}{\overline{V}}
\newcommand{\oR}{\overline{R}}
\newcommand{\oH}{\overline{H}}
\newcommand{\ol}{\overline{l}}
\newcommand{\E}{{\cal E}_d}
\newcommand{\el}{\ell}

\newcommand{\ej}{{e^{j2\pi f}}}

\newtheorem{theorem}{Theorem}


\newcommand{\yesindent}{\hspace*{\parindent}}   
\newtheorem{thmbody}{Theorem}
\newenvironment{thm}{
    \begin{singlespace} \begin{thmbody}
    }{
    \end{thmbody} \end{singlespace}
    }
\newtheorem{dfnbody}{Definition}
\newenvironment{dfn}{
    \begin{singlespace} \begin{dfnbody}
    }{
    \end{dfnbody} \end{singlespace}
    }
\newtheorem{corbody}{Corollary}
\newenvironment{cor}{
    \begin{singlespace} \begin{corbody}
    }{
    \end{corbody} \end{singlespace}
    }
\newtheorem{lemma}{Lemma}
\newtheorem{propbody}{Proposition}
\newenvironment{prop}{
    \begin{singlespace} \begin{propbody}
    }{
    \end{propbody} \end{singlespace}
    }
\newenvironment{example}{
    \begin{small} \begin{singlespace}{\it Example:}
    }{
    \end{singlespace} \end{small}
    }
\newcommand{\pderiv}[2]{\frac{ \partial {#1}}{ \partial {#2}}}
\newcommand{\overr}[2]{\left({\begin{array}{l}#1\\#2\end{array}}\right)}
 \newcommand{\Ddef}{\stackrel{\Delta}{=}}



\section{Introduction}
\label{intro0}
The {\em water-filling} solution for the quadratic
rate-distortion function $R(D)$ of a stationary Gaussian source is
given in terms of the spectrum of the source. Similarly, the
capacity $C$ of a power-constrained ISI
channel with Gaussian noise is given by a water-filling solution
relative to the effective noise spectrum. Both these formulas amount
to limiting values of mutual-information between vectors in the
frequency domain. In contrast, linear prediction along the time
domain can translate these vector mutual-information quantities into
scalar ones. Indeed, for capacity, Cioffi {\em et al}
\cite{MMSE-DFE} showed that $C$ is equal to the {\em scalar}
mutual-information over a slicer embedded in a decision-feedback
noise-prediction loop.

\begin{figure*}
\centering
\begin{picture}(0,0)%
\includegraphics{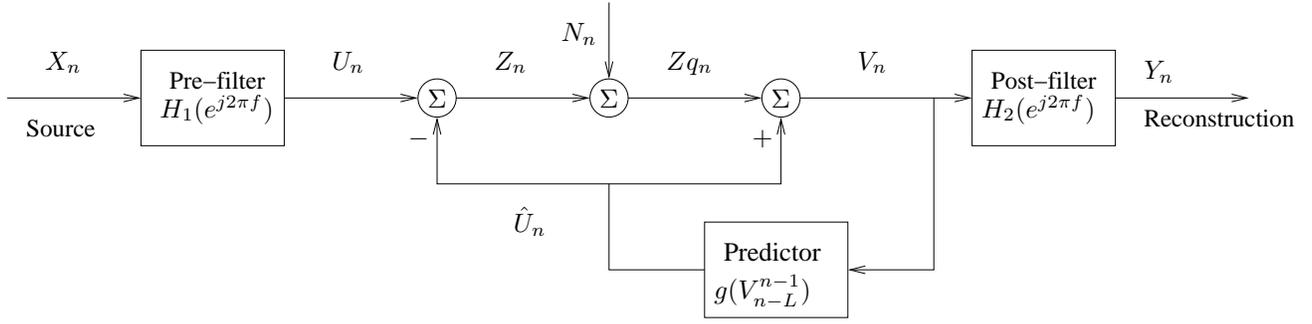}%
\end{picture}%
\setlength{\unitlength}{3158sp}%
\begingroup\makeatletter\ifx\SetFigFont\undefined%
\gdef\SetFigFont#1#2#3#4#5{%
  \reset@font\fontsize{#1}{#2pt}%
  \fontfamily{#3}\fontseries{#4}\fontshape{#5}%
  \selectfont}%
\fi\endgroup%
\begin{picture}(10275,2499)(514,-1873)
\put(3826,-211){\makebox(0,0)[lb]{\smash{{\SetFigFont{10}{12.0}{\rmdefault}{\mddefault}{\updefault}$\Sigma$}}}}
\put(826,
89){\makebox(0,0)[lb]{\smash{{\SetFigFont{10}{12.0}{\rmdefault}{\mddefault}{\updefault}$X_n$}}}}
\put(3076,
89){\makebox(0,0)[lb]{\smash{{\SetFigFont{10}{12.0}{\rmdefault}{\mddefault}{\updefault}$U_n$}}}}
\put(4351,
89){\makebox(0,0)[lb]{\smash{{\SetFigFont{10}{12.0}{\rmdefault}{\mddefault}{\updefault}$Z_n$}}}}
\put(5701,
89){\makebox(0,0)[lb]{\smash{{\SetFigFont{10}{12.0}{\rmdefault}{\mddefault}{\updefault}$Zq_n$}}}}
\put(7201,
89){\makebox(0,0)[lb]{\smash{{\SetFigFont{10}{12.0}{\rmdefault}{\mddefault}{\updefault}$V_n$}}}}
\put(9451,
14){\makebox(0,0)[lb]{\smash{{\SetFigFont{10}{12.0}{\rmdefault}{\mddefault}{\updefault}$Y_n$}}}}
\put(3676,-511){\makebox(0,0)[lb]{\smash{{\SetFigFont{10}{12.0}{\rmdefault}{\mddefault}{\updefault}$-$}}}}
\put(6376,-511){\makebox(0,0)[lb]{\smash{{\SetFigFont{10}{12.0}{\rmdefault}{\mddefault}{\updefault}$+$}}}}
\put(1726,-286){\makebox(0,0)[lb]{\smash{{\SetFigFont{10}{12.0}{\rmdefault}{\mddefault}{\updefault}$H_1(e^{j2\pi
f})$}}}}
\put(8176,-286){\makebox(0,0)[lb]{\smash{{\SetFigFont{10}{12.0}{\rmdefault}{\mddefault}{\updefault}$H_2(e^{j2\pi
f})$}}}}
\put(5176,-211){\makebox(0,0)[lb]{\smash{{\SetFigFont{10}{12.0}{\rmdefault}{\mddefault}{\updefault}$\Sigma$}}}}
\put(6526,-211){\makebox(0,0)[lb]{\smash{{\SetFigFont{10}{12.0}{\rmdefault}{\mddefault}{\updefault}$\Sigma$}}}}
\put(4876,314){\makebox(0,0)[lb]{\smash{{\SetFigFont{10}{12.0}{\rmdefault}{\mddefault}{\updefault}$N_n$}}}}
\put(4501,-1186){\makebox(0,0)[lb]{\smash{{\SetFigFont{10}{12.0}{\rmdefault}{\mddefault}{\updefault}$\hat
U_n$}}}}
\put(6151,-1411){\makebox(0,0)[lb]{\smash{{\SetFigFont{10}{12.0}{\rmdefault}{\mddefault}{\updefault}Predictor}}}}
\put(6076,-1711){\makebox(0,0)[lb]{\smash{{\SetFigFont{10}{12.0}{\rmdefault}{\mddefault}{\updefault}$g(V_{n-L}^{n-1})$}}}}
\end{picture}%
\caption{Predictive Test Channel.} \label{scheme_fig}
\end{figure*}

We show that a parallel result holds for the rate-distortion
function: $R(D)$ is equal to the {\em scalar} mutual-information
over an additive white Gaussian noise (AWGN) channel embedded in a
source prediction loop, as shown in \figref{scheme_fig}. This result
implies that $R(D)$ can essentially be realized in a sequential
manner (as will be clarified later), and it joins other observations
regarding the role of minimum mean-square error (MMSE) estimation in
successive encoding and decoding of Gaussian channels and sources
\cite{Forneyallerton04,ErezZamirAWGN,BergerMD}.


\subsection*{The Quadratic-Gaussian Rate-Distortion Function}
\label{intro}
\begin{figure}
\centering
\begin{picture}(0,0)%
\includegraphics{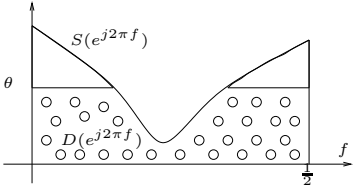}%
\end{picture}%
\setlength{\unitlength}{1579sp}%
\begingroup\makeatletter\ifx\SetFigFont\undefined%
\gdef\SetFigFont#1#2#3#4#5{%
  \reset@font\fontsize{#1}{#2pt}%
  \fontfamily{#3}\fontseries{#4}\fontshape{#5}%
  \selectfont}%
\fi\endgroup%
\begin{picture}(5538,2874)(1864,-3298)
\put(2926,-1111){\makebox(0,0)[lb]{\smash{{\SetFigFont{6}{7.2}{\rmdefault}{\mddefault}{\updefault}$S(e^{j2\pi
f})$}}}}
\put(2776,-2686){\makebox(0,0)[lb]{\smash{{\SetFigFont{6}{7.2}{\rmdefault}{\mddefault}{\updefault}$D(e^{j2\pi
f})$}}}}
\put(1876,-1786){\makebox(0,0)[lb]{\smash{{\SetFigFont{6}{7.2}{\rmdefault}{\mddefault}{\updefault}$\theta$}}}}
\put(7126,-2836){\makebox(0,0)[lb]{\smash{{\SetFigFont{6}{7.2}{\rmdefault}{\mddefault}{\updefault}$f$}}}}
\put(6526,-3211){\makebox(0,0)[lb]{\smash{{\SetFigFont{6}{7.2}{\rmdefault}{\mddefault}{\updefault}$\frac{1}{2}$}}}}
\end{picture}%
\caption{The water filling solution.} \label{water_filling_fig}
\end{figure}

The rate-distortion function (RDF) of a stationary
source with memory is given as a limit of normalized mutual
information associated with vectors of source samples.
For a real valued source
$\{X_n\} = \ldots, X_{-2}, X_{-1}, X_0, X_1, X_2,\ldots$,
and expected mean-squared distortion level $D$,
the RDF can be written as,
\cite{Berger71},
\[
R(D) = \lim_{n \rightarrow \infty} \frac{1}{n}
\inf I(X_1,\ldots,X_n; Y_1,\ldots,Y_n)
\]
where the infimum is over all channels $\bX \rightarrow \bY$
such that $\frac{1}{n} \|\bY-\bX\|^2 \leq D$.
A channel which realizes this infimum is called
an {\em optimum test-channel}.
When the source is zero-mean Gaussian, the RDF takes an explicit form
in the frequency domain in terms of the power-spectrum
\[
S(e^{j 2 \pi f}) = \sum_k R[k] e^{-j k 2 \pi f}   ,  \ \ \  -1/2 < f < 1/2 ,
\]
where
$R[k] = E\{ X_n X_{n+k} \}$
is the auto-correlation function of the source.
The water filling solution,
illustrated in Figure~\ref{water_filling_fig},
gives a parametric formula for
the Gaussian RDF in terms of a parameter $\theta$
\cite{Gallager68,Berger71,CoverBook}:
\begin{eqnarray}
\label{qgrdf}
\nonumber
R(D) &=&
\int_{-1/2}^{1/2} \frac{1}{2} \log \left( \frac{S(e^{j 2 \pi f})}{D(e^{j 2 \pi f})} \right) df  \\
&=&  \int_{f: S(e^{j 2 \pi f}) > \theta} \frac{1}{2} \log \left( \frac{S(e^{j 2 \pi f})}{\theta} \right) df
\end{eqnarray}
where the {\em distortion spectrum} is given by
\beq{water_filling_distortion}
D(e^{j 2 \pi f}) = \left\{
\begin{array}{ll}
\theta, & \mbox{if $S(e^{j 2 \pi f}) > \theta$} \\
S(e^{j 2 \pi f}), & \mbox{otherwise,}
\end{array}
\right.
\eeq
and where we choose the {\em water level} $\theta$ so that the total distortion
is $D$:
\beq{total_distortion}
D = \int_{-1/2}^{1/2} D(e^{j 2 \pi f}) df  .
\eeq

In the special case of a memoryless (white) Gaussian source
$\sim N(0, \sigma^2)$, the power-spectrum is flat $S(e^{j 2 \pi f})=\sigma^2$,
so $\theta = D$
and the RDF is simplified to
\beq{memoryless}
\frac{1}{2} \log \left( \frac{\sigma^2}{D} \right)
\ , \ \ \ 0 < D \leq \sigma^2 .
\eeq
The optimum test-channel can be written in this case in
a {\em backward} additive-noise form:  $X = Y + N$,
with $N \sim N(0,D)$,
or in a {\em forward} linear additive-noise form:
\[
Y = \beta ( \alpha X + N)
\]
with $\alpha = \beta = \sqrt{1-D/\sigma^2}$
and $N \sim N(0,D)$.
In the general stationary case, the forward channel realization of
the Gaussian RDF has several equivalent forms \cite[Sec.
9.7]{Gallager68}, \cite[Sec. 4.5]{Berger71}. The one which is more
useful for our purpose replaces $\alpha$ and $\beta$ above by linear
time-invariant filters, while keeping the noise $N$ as AWGN
\cite{FederZamir94}: \beq{forward} Y_n = h_{2,n} * ( h_{1,n} * X_n +
N_n ) \eeq where $N_n \sim N(0,\theta)$ is AWGN with $\theta =
\theta(D) =$ the water level, $*$ denotes convolution, and $h_{1,n}$
and $h_{2,n}$ are the impulse responses of a suitable pre-filter and
post-filter, respectively.
See (\ref{prefilter})-(\ref{postfilter}) in the next section.

If we take a discrete approximation of  (\ref{qgrdf}),
\beq{qgrdf_aprx}
\sum_i  \frac{1}{2} \log \left( \frac{S(e^{j 2 \pi f_i})}{D(e^{j 2 \pi f_i})} \right) ,
\eeq
then each component has the memoryless form of (\ref{memoryless}).
Hence, we can think of the frequency domain formula (\ref{qgrdf}) as
an encoding of {\em parallel} (independent) Gaussian sources, where source $i$ is a
memoryless Gaussian source
$X_{i} \sim  N(0, S(e^{j 2 \pi f_i}))$
encoded at distortion level
$D(e^{j 2 \pi f_i})$; see \cite{CoverBook}.
Indeed, practical frequency domain source coding schemes such as
Transform Coding and Sub-band Coding \cite{GibsonBerger} get close
to the RDF of a stationary Gaussian source using an ``array'' of
parallel {\em scalar} quantizers.


\subsection*{Rate-Distortion and Prediction}
Our main result is a predictive channel realization for the
quadratic-Gaussian RDF (\ref{qgrdf}), which can be viewed as
the time-domain counterpart of the frequency domain
formulation above.
The notions of {\em entropy-power} and {\em Shannon lower bound}
(SLB)
provide a simple relation between the Gaussian RDF and prediction,
and motivate our result.
Recall that the entropy-power is the variance of a {\em white}
Gaussian process having the same entropy-rate as the source
\cite{CoverBook};
for a zero-mean Gaussian source with power-spectrum $S(e^{j 2 \pi f})$,
the entropy-power is given by
\beq{entropy_power}
P_e(X) = \exp \left( \int_{-1/2}^{1/2} \log \left( S(e^{j 2 \pi f}) \right) df
\right) .
\eeq

In the context of Wiener's spectral-factorization theory,
the entropy-power quantifies the MMSE
in one-step linear prediction of a Gaussian source from its infinite past
\cite{Berger71}:
\beq{one_step_prediction}
P_e(X) = \inf_{\{a_i\}}
E \left( X_n  -  \sum_{i=1}^\infty a_i X_{n-i}  \right)^2 .
\eeq
The error process associated with the infinite-order optimum predictor,
\beq{innovation}
Z_n = X_n  -  \sum_{i=1}^\infty a_i X_{n-i} ,
\eeq
is called the {\em innovation process}.
The {\em orthogonality principle} of MMSE estimation implies that
the innovation process has zero mean and is {\em white};
in the Gaussian case un-correlation implies independence, so
\beq{innovation-var}
Z_n \sim {\cal N}(0, P_e(X))
\eeq
is a {\em memoryless} process.
See, e.g., \cite{Forneyallerton04}.

>From an information theoretic perspective,
the entropy-power plays a role in the SLB:
\beq{SLB}
R(D) \geq
\frac{1}{2} \log \left( \frac{P_e(X)}{D} \right) .
\eeq
Equality in the SLB holds if the distortion level is smaller than or equal to the
lowest value of the power spectrum:
$D \leq S_{min} \Ddef \min_f S(e^{j 2 \pi f})$,
in which case  $D(e^{j 2 \pi f})=\theta=D$
\cite{Berger71}.
%
It follows that for distortion levels below $S_{min}$ the RDF of
a Gaussian source with memory is equal to the RDF of its
memoryless innovation process $Z_n$:
\beq{RDinnovation}
R(D) = R_Z(D) = \frac{1}{2}
\log \left( \frac{ \sigma_Z^2 }{D} \right) , \ \ D \leq S_{min},
\eeq
where $\sigma_Z^2 = P_e(X)$.

\begin{figure}
\centering
\begin{picture}(0,0)%
\includegraphics{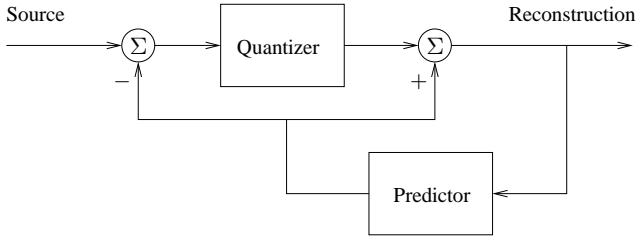}%
\end{picture}%
\setlength{\unitlength}{2723sp}%
\begingroup\makeatletter\ifx\SetFigFont\undefined%
\gdef\SetFigFont#1#2#3#4#5{%
  \reset@font\fontsize{#1}{#2pt}%
  \fontfamily{#3}\fontseries{#4}\fontshape{#5}%
  \selectfont}%
\fi\endgroup%
\begin{picture}(5771,2124)(2689,-1873)
\put(3826,-211){\makebox(0,0)[lb]{\smash{{\SetFigFont{9}{10.8}{\rmdefault}{\mddefault}{\updefault}$\Sigma$}}}}
\put(3676,-511){\makebox(0,0)[lb]{\smash{{\SetFigFont{8}{9.6}{\rmdefault}{\mddefault}{\updefault}$-$}}}}
\put(6376,-511){\makebox(0,0)[lb]{\smash{{\SetFigFont{8}{9.6}{\rmdefault}{\mddefault}{\updefault}$+$}}}}
\put(6526,-211){\makebox(0,0)[lb]{\smash{{\SetFigFont{9}{10.8}{\rmdefault}{\mddefault}{\updefault}$\Sigma$}}}}
\put(4801,-211){\makebox(0,0)[lb]{\smash{{\SetFigFont{8}{9.6}{\rmdefault}{\mddefault}{\updefault}Quantizer}}}}
\put(2701,
89){\makebox(0,0)[lb]{\smash{{\SetFigFont{8}{9.6}{\rmdefault}{\mddefault}{\updefault}Source}}}}
\put(6226,-1561){\makebox(0,0)[lb]{\smash{{\SetFigFont{8}{9.6}{\rmdefault}{\mddefault}{\updefault}Predictor}}}}
\put(7276,
89){\makebox(0,0)[lb]{\smash{{\SetFigFont{8}{9.6}{\rmdefault}{\mddefault}{\updefault}Reconstruction}}}}
\end{picture}%
\caption{DPCM Quantiztion Scheme.} \label{DPCM_fig}
\end{figure}

We shall see later in Section~II how identity (\ref{RDinnovation}) translates into
a predictive test-channel, which can realize the RDF
not only for small but for {\em all} distortion levels.
%
This test channel
is motivated by the sequential structure
of Differential Pulse Code Modulation (DPCM)
\cite{Jayant84,GibsonBerger}.
The goal of DPCM is to translate the encoding of dependent source
samples into a series of {\em independent} encodings.
The task of removing the time dependence is
achieved by (linear) prediction:
at each time instant the incoming source sample is predicted
from previously encoded samples, the prediction error
is encoded by a scalar quantizer and added to the predicted value
to form the new reconstruction.
See \figref{DPCM_fig}.

A negative result along this direction was recently given by Kim and
Berger \cite{BergerD*PCM}. They showed that the RDF of an
auto-regressive (AR) Gaussian process cannot be achieved by directly
encoding its innovation process. This can be viewed as {\em
open-loop} prediction, because the innovation process is extracted
from the clean source rather than from the quantized source
\cite{Jayant84,GershoGray}. Here we give a positive result, showing
that the RDF can be achieved if we embed the quantizer inside the
prediction loop, i.e., by {\em closed-loop} prediction as done in
DPCM. The RDF-achieving system consists of pre- and post-filters,
and an AWGN channel embedded in a source prediction loop. As we
show, the {\em scalar} (un-conditioned) mutual information over this
inner AWGN channel is equal to the RDF.

After presenting and proving our main result in Sections~II and~III,
respectively, we discuss its characteristics and operational
implications. Section~IV discusses the spectral features of the
solution. Section~V relates the solution to vector-quantized DPCM of
parallel sources. Section~VI shows an implementation by Entropy
Coded Dithered Quantization (ECDQ), while extending the ECDQ rate
formula \cite{FederZamir} to the case of a system with feedback.
Finally, in Section~VII we relate prediction in source coding to
prediction for channel equalization and to recent observations by
Forney \cite{Forneyallerton04}.  As in \cite{Forneyallerton04}, our
analysis is based on the properties of information measures; the
only result we need from Wiener's estimation theory is the
orthogonality principle.


\vspace{1cm}

\section{Main Result}
\label{main_sec}
Consider the system in \figref{scheme_fig}, which consists of
three basic blocks:
a pre-filter $H_1(e^{j 2 \pi f})$, a noisy channel embedded
in a closed loop, and a post-filter $H_2(e^{j 2 \pi f})$,
where $H(e^{j 2 \pi f})$ denotes the frequency response
of a filter with impulse response $h_n$,
\[
H(e^{j 2 \pi f}) = \sum_n h_n e^{-j n 2 \pi f}   ,  \ \ \  -1/2 < f < 1/2 .
\]
The system parameters are derived from the water-filling
solution (\ref{qgrdf})-(\ref{water_filling_distortion}),
and depend on the source spectrum $S(e^{j 2 \pi f})$
and the distortion level $D$.
The source samples
$\{X_n\}$ are passed through a pre-filter, whose phase is
arbitrary and its absolute squared frequency response is given by
\begin{eqnarray} \label{prefilter}
|H_1(e^{j 2 \pi f})|^2 &=& 1 - \frac{D(e^{j 2 \pi f})}{S(e^{j 2 \pi f})}
\end{eqnarray}
where $\frac{0}{0}$ is taken as 1.
The pre-filter output, denoted $U_n$, is fed to the central block which
generates a process $V_n$ according to the following recursion
equations:
\begin{eqnarray}
\label{recursive_test_channel1}
\hU_n &=&  g(V_{n-1}, V_{n-2}, \ldots, V_{n-L}) \\
Z_n &=& U_n - \hU_n  \\
\label{coreAWGN}
Zq_n &=& Z_n + N_n  \\
\label{recursive_test_channel2}
V_n &=& \hU_n + Zq_n
\end{eqnarray}
%
where     
$N_n \sim \cN(0,\theta)$ is a zero-mean white Gaussian noise,
independent of the input process $\{U_n\}$,
whose variance is equal to the water level $\theta = \theta(D)$; and
$g(\cdot)$ is some prediction function for the input $U_n$ given the
$L$ past samples of the output process $(V_{n-1}, V_{n-2}, \ldots,
V_{n-L})$. \footnote {No initial condition on $V_n$ is needed as we
assume a two-sided input process $X_n$, and the system is stable.}
Finally, the post-filter frequency response is the complex conjugate
of the frequency response of the pre-filter, \beq{postfilter}
H_2(e^{j 2 \pi f}) = H^*_1(e^{j 2 \pi f}) . \eeq Equivalently, the
impulse response of the post-filter is the reflection of the impulse
response of the pre-filter: \beq{postfilterA} h_{2,n} = h_{1,-n} \ \
. \eeq See a comment regarding {\em causality} in the end of the
section.

\begin{figure}[!t]
\centering
\begin{picture}(0,0)%
\includegraphics{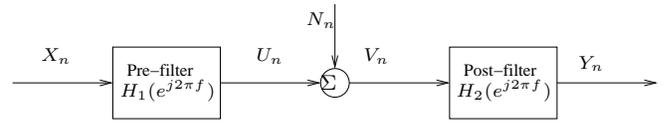}%
\end{picture}%
\setlength{\unitlength}{2368sp}%
\begingroup\makeatletter\ifx\SetFigFont\undefined%
\gdef\SetFigFont#1#2#3#4#5{%
  \reset@font\fontsize{#1}{#2pt}%
  \fontfamily{#3}\fontseries{#4}\fontshape{#5}%
  \selectfont}%
\fi\endgroup%
\begin{picture}(6774,1224)(3514,-523)
\put(7201,
89){\makebox(0,0)[lb]{\smash{\SetFigFont{7}{8.4}{\rmdefault}{\mddefault}{\updefault}$V_n$}}}
\put(9451,
14){\makebox(0,0)[lb]{\smash{\SetFigFont{7}{8.4}{\rmdefault}{\mddefault}{\updefault}$Y_n$}}}
\put(6751,-211){\makebox(0,0)[lb]{\smash{\SetFigFont{8}{9.6}{\rmdefault}{\mddefault}{\updefault}$\Sigma$}}}
\put(6601,464){\makebox(0,0)[lb]{\smash{\SetFigFont{7}{8.4}{\rmdefault}{\mddefault}{\updefault}$N_n$}}}
\put(6076,
89){\makebox(0,0)[lb]{\smash{\SetFigFont{7}{8.4}{\rmdefault}{\mddefault}{\updefault}$U_n$}}}
\put(3826,
89){\makebox(0,0)[lb]{\smash{\SetFigFont{7}{8.4}{\rmdefault}{\mddefault}{\updefault}$X_n$}}}
\put(4651,-286){\makebox(0,0)[lb]{\smash{\SetFigFont{7}{8.4}{\rmdefault}{\mddefault}{\updefault}$H_1(e^{j2\pi
f})$}}}
\put(8176,-286){\makebox(0,0)[lb]{\smash{\SetFigFont{7}{8.4}{\rmdefault}{\mddefault}{\updefault}$H_2(e^{j2\pi
f})$}}}
\end{picture}
\caption{Equivalent Channel.} \label{equivalent_fig}
\end{figure}

The block from $U_n$ to $V_n$ is equivalent to the configuration of
DPCM,
\cite{Jayant84,GibsonBerger},
with the DPCM quantizer replaced by the additive Gaussian noise channel
$Zq_n = Z_n + N_n$.
In particular, the recursion equations
(\ref{recursive_test_channel1})-(\ref{recursive_test_channel2})
imply that this block satisfies the well known ``DPCM error identity'',
\cite{Jayant84},
\beq{DPCMerror}
V_n = U_n + (Zq_n - Z_n) = U_n + N_n  .
\eeq
That is, the output $V_n$ is a noisy version of the input $U_n$
via the AWGN channel $V_n = U_n + N_n$.
Thus, the system of \figref{scheme_fig}
is equivalent to the system depicted in \figref{equivalent_fig},
which corresponds to the forward channel realization
(\ref{forward})
of the quadratic-Gaussian RDF.

In DPCM the prediction function $g$ is linear:
\beq{linear_predictor} g(V_{n-1}, \ldots, V_{n-L}) = \sum_{i=1}^L
a_i V_{n-i} \eeq where $a_1,\ldots,a_L$ are chosen to minimize the
mean-squared prediction error:
\beq{prediction_error}
\sigma^2_L = \min_{a_i}
E \left( U_n  - \sum_{i=1}^L a_i V_{n-i} \right)^2 .
\eeq
Because $V_n$ is the result of passing $U_n$ through
an AWGN channel, we call that ``noisy prediction''.
If $\{U_n\}$ and $\{V_n\}$ are jointly Gaussian, then the best
predictor of any order is linear, so $\sigma_L^2$
is also the MMSE in estimating $U_n$ from
the vector $(V_{n-1}, \ldots, V_{n-L})$.
Clearly, this MMSE is non-increasing with the prediction order $L$,
and as $L$ goes to infinity it converges to
\beq{prediction_error_infinite}
\sigma^2_\infty = \lim_{L \rightarrow \infty} \sigma^2_L,
\eeq
the optimum infinite order prediction
error in $U_n$ given the past
\beq{past} V_n^- \Ddef \{
V_{n-1}, V_{n-2}, \ldots \} .
\eeq
We shall see later in Section~\ref{properties_sec} that
$\sigma^2_\infty = P_e(V) - \theta$.
We further elaborate on the
relationship with DPCM in Section~\ref{DPCM_sec}.
We now state our main result.

\vspace{.3in}

\begin{theorem}
\label{thm1} \textbf{(Predictive test channel)}
For any stationary source with power spectrum $S(e^{j 2 \pi f})$
and distortion level $D$,
the system of \figref{scheme_fig},
with the pre-filter (\ref{prefilter}) and the post-filter
(\ref{postfilter}),
satisfies \beq{main0} E(Y_n - X_n)^2 = D . \eeq Furthermore, if the
source $X_n$ is Gaussian and  $g = g(V_n^-)$ achieves the optimum
infinite order prediction error $\sigma_\infty^2$
(\ref{prediction_error_infinite}),
then
\begin{equation}
\label{main} I(Z_n; Z_n + N_n) = \frac{1}{2} \log( 1 +
\frac{\sigma^2_\infty}{\theta} ) = R(D) ,
\end{equation}
where the left hand side is the {\em scalar} mutual information over
the channel (\ref{coreAWGN}).
\end{theorem}

\vspace{.3in}

The proof is given in \secref{proof_sec}.
%
The result above is in sharp contrast to the classical realization of the RDF (\ref{forward}),
which involves mutual information {\em rate} over
a test-channel with {\em memory}.
In a sense, the core of the encoding process in the system of
\figref{scheme_fig} amounts to
a {\em memoryless AWGN test-channel}
(although, as we discuss in the sequel, the channel (\ref{coreAWGN})
is not quite memoryless nor additive).
>From a practical perspective,
this system provides a bridge between
DPCM and rate-distortion theory
for a general distortion level $D>0$.

Another interesting feature of the system is the relationship
between the prediction error process $Z_n$ and the original
process $X_n$.
If $X_n$ is an auto-regressive (AR) process,
then in the limit of small distortion ($D \rightarrow 0$),
$Z_n$ is roughly its innovation process (\ref{innovation}).
Hence,
unlike in open-loop prediction \cite{BergerD*PCM}, encoding the
innovations in a closed-loop system is optimal in the limit of
high-resolution encoding. We shall return to this point, as well as
discuss the case of general resolution, in
\secref{properties_sec}.

Finally,
we note that while the central block of the system is sequential and hence causal,
the pre- and post-filters are non-causal and therefore their realization
in practice requires {\em delay}.
Specifically, since by (\ref{postfilterA})
$h_{2,n} = h_{1,-n}$,
if one of the filters is causal then the other must be anti-causal.
Often the filter's response is infinite, hence the required
delay is infinite as well.
{
Of course, one can approximate the desired spectrum
(in $L_2$ sense and hence also in rate-distortion sense)
to any degree using filters of sufficiently large but finite delay $\delta$,
so the system distortion is actually measured between
$Y_n$ and $X_{n- \delta}$.
In this sense, Theorem~\ref{thm1} holds in general in the limit as the
system delay $\delta$ goes to infinity.

If we insist on a system with {\em causal} reconstruction
($\delta=0$), then we cannot realize the pre- and post-filters
(\ref{prefilter}) and (\ref{postfilter}), and some loss in
performance must be paid. Nevertheless, if the source spectrum is
bounded from below by a positive constant, then it can be seen from
(\ref{prefilter}) that in the limit of small distortion ($D
\rightarrow 0$) the filters can be omitted, i.e., $H_1 = H_2 = 1$
for all $f$. Hence, a causal system (the central block in
\figref{scheme_fig}) is asymptotically optimal at ``high
resolution'' conditions. Furthermore, the redundancy of an AWGN
channel above the RDF is at most $0.5$ bit per source sample for
{\em any} source and at {\em any} resolution; see, e.g.,
\cite{FederZamir}. It thus follows from Lemma~\ref{lem1} below
(which directly characterizes the information rate of the central
block of \figref{scheme_fig}), that a causal system (the system of
\figref{scheme_fig} without the filters) loses at most 0.5 bit at
any resolution.

These observations shed some light on the
``cost of causality'' in encoding stationary Gaussian sources
\cite{LinderZamirCausal}.
It is an open question, though, whether a redundancy better than 0.5 bit can be guaranteed
when using {\em causal} pre and post filters in the system of \figref{scheme_fig}.





\vspace{1cm}

\section{Proof of Main Result}
\label{proof_sec}

We start with Lemma~\ref{lem1} below, which shows an identity
between the mutual information rate over the central block
of \figref{scheme_fig}
and the scalar mutual information (\ref{main}).
This identity holds regardless of the pre- and post-filters,
and only assumes optimum infinite order prediction in the feedback loop.

Let
\beq{mutual_info_rate}
\oI(\{U_n\}; \{V_n\})
= \lim_{n \rightarrow \infty} \frac{1}{n}
I(U_1,\ldots,U_n; V_1,\ldots,V_n)
\eeq
denote mutual information-rate between
jointly stationary sources $\{U_n\}$ and $\{V_n\}$,
whenever the limit exists.

\vspace{.3in}

\begin{lemma}
\label{lem1}
For any stationary Gaussian process $\{U_n\}$ in
\figref{scheme_fig}, if $\hat{U}_n$ is the optimum infinite order
predictor of $U_n$ from $V_n^-$
(so the variance of $Z_n$ is $\sigma_\infty^2$ as defined
in (\ref{prediction_error_infinite})), then \beq{coreA} \oI(\{U_n\};
\{V_n\}) =  I(Z_n; Z_n + N_n) . \eeq
\end{lemma}

\vspace{.3in}

\begin{proof}
For any finite order predictor $g(V_{n-L}^{n-1})$ we can write
\begin{eqnarray}
\nonumber
I(\{U_n\}; V_i |  V_{i-L}^{i-1})
& = &
I(\{U_n\}, U_i - \hU_i^{(L)} ; V_i - \hU_i^{(L)} |  V_{i-L}^{i-1})
\\
\nonumber
& = &
I(\{U_n\}, Z_i^{(L)}; Z_i^{(L)} + N_i|  V_{i-L}^{i-1})
\\
\label{markov}
& = &
I(Z_i^{(L)}; Z_i^{(L)} + N_i|  V_{i-L}^{i-1})
\\
\label{ortho}
& = &
I(Z_i^{(L)}; Z_i^{(L)} + N_i)
\end{eqnarray}
where $\hU_i^{(L)} = g(V_{i-L}^{i-1})$ is the $L$-th order predictor
output at time $i$, and $Z_i^{(L)}$ is the prediction error. The
first equality above follows since manipulating the condition does
not affect the conditional mutual information; the second equality
follows from the definition of $Z_i^{(L)}$; (\ref{markov}) follows
since $N_i$ is independent of $(\{U_n\}, V_i^-)$ and therefore
$$(Z_i^{(L)}+N_i)  \longleftrightarrow
(Z_i^{(L)}, V_{i-L}^{i-1})  \longleftrightarrow \{U_n\}$$
form a Markov chain;
and (\ref{ortho}) follows from two facts:
first, since $N_i$ is independent of $\{U_i\}$ and previous $N_i$'s,
it is also independent of the pair $(Z_i^{(L)}, V_{i-L}^{i-1})$
by the recursive structure of the system;
second, we assume optimum (MMSE) prediction,
hence the orthogonality principle implies that
the prediction error $Z_i^{(L)}$ is orthogonal to
the measurements  $V_{i-L}^{i-1}$, so by Gaussianity they are also independent,
and hence by the two facts we have that
$V_{i-L}^{i-1}$ is independent of the pair  $(Z_i^{(L)}, N_i)$.
Since by (\ref{prediction_error})
the variance of the $L$-th order prediction error $Z_i^{(L)}$
is $\sigma_L^2$,
while the variance of the noise $N_i$ is $\theta$,
we thus obtained from (\ref{ortho})
\beq{coreB}
I(\{U_n\}; V_i |  V_{i-L}^{i-1}) = \frac{1}{2} \log
\Bigl( 1 + \frac{\sigma_L^2}{\theta}  \Bigr)  .
\eeq
This implies in
the limit as $L \rightarrow \infty$
\begin{eqnarray}
I(\{U_n\}; V_i | V_i^-)
&=& \frac{1}{2} \log \Bigl( 1 + \frac{\sigma_\infty^2}{\theta}
\Bigr)
\\
\label{coreC}
&=& I(Z_n; Z_n + N_n).
\end{eqnarray}
Note that by stationarity,  $I(\{U_n\}; V_i |  V_i^-)$ is
independent of $i$.
Thus,
\[
I(\{U_n\}; V_1) + I(\{U_n\}; V_2 |  V_1) + \ldots +
I(\{U_n\}; V_i |  V_{1}^{i-1})
\]
normalized by $1/i$
converges as $i \rightarrow \infty$ to $I(\{U_n\}; V_i |  V_i^-)$.
By the definition of mutual information rate (\ref{mutual_info_rate})
and by the chain rule for mutual information \cite{CoverBook},
this implies that the left hand side of (\ref{coreA})
is equal to
\beq{mi-rateB}
\oI(\{U_n\}; \{V_n\}) = I(\{U_n\}; V_i |  V_i^-) .
\eeq
Combining (\ref{coreC}) and (\ref{mi-rateB}) the lemma is proved.
\end{proof}

\vspace{.3in}

Theorem~\ref{thm1} is a simple consequence
of Lemma~\ref{lem1} above and the forward channel realization
of the RDF.
As discussed in the previous section,
the DPCM error identity (\ref{DPCMerror}) implies that
the entire system of \figref{scheme_fig}
is equivalent to the system depicted in \figref{equivalent_fig},
consisting of a pre-filter (\ref{prefilter}), an AWGN channel with
noise variance $\theta$, and a post-filter (\ref{postfilter}).
This is also the forward channel realization (\ref{forward}) of the RDF
\cite{Gallager68,Berger71,FederZamir94}.
In particular, as simple spectral analysis shows, the power
spectrum of the overall error process $Y_n-X_n$ is equal
to the water filling distortion spectrum $D(e^{j 2 \pi f})$
in (\ref{water_filling_distortion}).
Hence, by (\ref{total_distortion}) the total distortion is $D$,
and (\ref{main0}) follows.

We turn to prove the second part of the theorem
(equation~(\ref{main}) ).
Since the system of \figref{equivalent_fig}
is equivalent to the forward channel realization (\ref{forward}) of
the RDF of $\{X_n\}$,
we have
\beq{forwardB}
\oI(\{X_n\}; \{Y_n\}) = R(D)
\eeq
where $\oI$ denotes mutual information-rate (\ref{mutual_info_rate}).
Since $\{U_n\}$ is a function of $\{X_n\}$, and since
the post-filter $H_2$ is invertible within the pass-band of
the pre-filter $H_1$,
we also have
\beq{forwardC}
\oI(\{X_n\}; \{Y_n\}) = \oI(\{U_n\}; \{V_n\})  .
\eeq
The theorem now follows by combining (\ref{forwardC}),
(\ref{forwardB}) and  Lemma~\ref{lem1}.


An alternative proof of Theorem~\ref{thm1}, based only on spectral considerations,
is given in the end of the next section.

\vspace{1cm}


\section{Properties of the Predictive Test-Channel}
\label{properties_sec}
The following observations shed light on the behavior of the test
channel of \figref{scheme_fig}.

\textbf{Prediction in the high resolution regime.}
If the power-spectrum $S(e^{j 2 \pi f})$
is everywhere positive (e.g., if $\{X_n\}$
can be represented as an AR process), then in the limit of small
distortion $D \rightarrow 0$,
the pre- and post-filters
(\ref{prefilter}), (\ref{postfilter}) converge to
all-pass filters,
and the power spectrum of $U_n$ becomes the power spectrum
of the source $X_n$.
Furthermore,  noisy prediction of $U_n$ (from the ``noisy past''
$V_n^{-}$,
where $V_n=U_n+N_n$)
becomes equivalent to clean prediction of $U_n$
from its own past $U_n^{-}$.
Hence, in this limit the prediction error $Z_n$ is equivalent
to the innovation process of $X_n$ (\ref{innovation}).
In particular,
$Z_n$ is an i.i.d. process whose variance is $P_e(X)$ = the entropy-power
of the source (\ref{entropy_power}).

\textbf{Prediction in the general case.} Interestingly, for
general distortion $D>0$, the prediction error $Z_n$ is \emph{not
white}, as the noisiness of the past does not allow the predictor
$g$ to remove all the source memory.   Nevertheless, the noisy
version of the prediction error $Zq_n = Z_n + N_n$ is white for
every $D>0$, because it amounts to predicting $V_n$ from its {\em
own} infinite past:
since $N_n$ has zero-mean and is white (and therefore independent
of the past), $\hat{U}_n$ that minimizes the prediction error of $U_n$
is also the optimal predictor for
$V_n=U_n+N_n$.
In particular, in view of (\ref{one_step_prediction}) and (\ref{innovation-var}),
we have
\begin{equation}
\label{tricky}
Zq_n \sim {\cal N}(0, P_e(V))
\end{equation}
where $P_e(V)$ is the entropy-power of the process $V_n$.
And since $Zq_n$ is the independent sum of $Z_n$ and $N_n$, we
also have the relation
\[
P_e(V) = \sigma_\infty^2 + \theta
\]
where $\sigma_\infty^2$ is the variance of $Z_n$ (\ref{prediction_error_infinite})
and $\theta$ is the variance of $N_n$.


\textbf{Sequential Additivity.}  
The whiteness of $Zq_n$ might seem at first a contradiction,
because $Zq_n$ is the sum of a non-white process, $Z_n$, and a white
process $N_n$;
nevertheless,
$\{Z_n\}$ and $\{N_n\}$ are \emph{not} independent,
because $Z_n$ depends on past values of
$N_n$ through the feedback loop and the past of $V_n$.
Thus, the channel $Zq_n = Z_n + N_n$ is not quite additive
but ``sequentially additive'':
each new noise sample is independent of the present and
the past but not necessarily of the future.
In particular, this channel satisfies:
\beq{directedI}
I(Z_n; Z_n+N_n | Z_1+N_1,...,Z_{n-1}+N_{n-1}) =
I(Z_n; Z_n+N_n)\ ,
\eeq
so by the chain rule for mutual information
\[ \bar{I}(\{Z_n\}; \{Z_n+N_n\}) >
I(Z_n; Z_n+N_n) \ . \]
Later in Section~\ref{ECDQ_sec} we rewrite (\ref{directedI}) in terms of
directed mutual information.


\textbf{The channel when the SLB is tight.} As long
as $D$ is smaller than the lowest point of the source power spectrum $S_{min}$,
we have
$D(e^{j 2 \pi f})=\theta=D$ in \eqref{qgrdf},
and the quadratic-Gaussian
RDF coincides with the SLB (\ref{SLB}).
In this case, the following properties hold for the predictive test channel:
\begin{itemize}
    \item{The power spectra of $U_n$ and $Y_n$ are the same and are equal to
$S(e^{j 2 \pi f})-D$.}
    \item{The power spectrum of $V_n$ is equal to the power spectrum of
    the source $S(e^{j 2 \pi f})$.}
    \item{
    The variance of $Zq_n$ is equal to the entropy-power 
    of $V_n$ by (\ref{tricky}), which is equal to $P_e(X)$.}
    \item{As a consequence we have
    \begin{eqnarray*}
    I(Z_n; Z_n+N_n) &=& h(Zq_n) - h(N_n)  \\
    &=& h\Bigl( \cN(0,P_e(V)) \Bigr) - h\Bigl( \cN(0,D) \Bigr) \\
    &=&  \frac{1}{2}  \log\Bigl( \frac{P_e(X)}{D} \Bigr)
    \end{eqnarray*}
    which is indeed the SLB (\ref{SLB}).
    }
    \end{itemize}
As discussed in the Introduction, the SLB is also the RDF of the
innovation process (\ref{RDinnovation}), i.e.,
the conditional RDF of the source $X_n$ given its infinite
{\em clean} past $X_n^-$.


\textbf{An alternative derivation of Theorem~\ref{thm1} in the spectral domain.}
For a general $D$, we can use (\ref{tricky}) and the equivalent channel of
\figref{equivalent_fig}
to re-derive the scalar mutual information - RDF identity (\ref{main}).
Note that for any $D$ the power spectrum of $U_n$
and $Y_n$ is equal to $\max\{ 0, S(e^{j 2 \pi f})-\theta \}$,
where $\theta = \theta(D)$ is the water-level.
Thus the power spectrum of $V_n=U_n + N_n$ is given by
$\max\{ \theta, S(e^{j 2 \pi f}) \}$.
Since as discussed above the variance of $Zq_n=Z_n+N_n$ is given
by the entropy power of the process $V_n$, we have
\begin{eqnarray}
\nonumber
I(Z_n; Z_n+N_n) &=&
 \frac{1}{2}  \log \Bigl(
\frac{P_e( \max\{ \theta, S(e^{j 2 \pi f}) \} ) }{\theta} \Bigr) \\
\nonumber
&=& R(D)
\end{eqnarray}
where $P_e(\cdot)$ as a function of the spectrum is given in (\ref{entropy_power}),
and the second equality follows from (\ref{qgrdf}).



\vspace{1cm}

\section{Vector-Quantized DPCM  and  D$^*$PCM}
\label{DPCM_sec}
As mentioned earlier, the structure of the central block of the
channel of \figref{scheme_fig} is of a DPCM encoder,
with the scalar quantizer
replaced by the AWGN channel $Zq_n=Z_n+N_n$. However, if we wish to
implement the additive noise by a quantizer whose rate is the mutual
information $I(Z_n; Z_n + N_n)$, we must use \emph{vector}
quantization (VQ).
Indeed, while scalar quantization noise is approximately uniform
over intervals, good high dimensional lattices
generate near Gaussian quantization noise \cite{LQN}. Yet, how can we
combine VQ and DPCM without violating the sequential nature of the
system? In particular, the quantized sample $Zq_n$ must be available
to generate $V_n$,
before the system can predict $U_{n+1}$ and generate $Z_{n+1}$.

\begin{figure}[!t]
\centering
\begin{picture}(0,0)%
\includegraphics{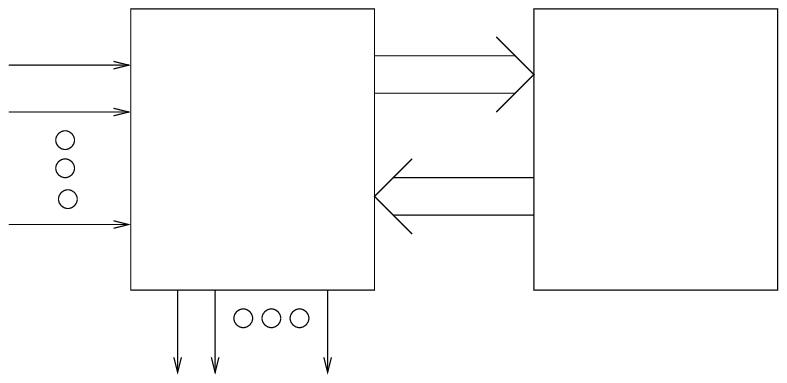}%
\end{picture}%
\setlength{\unitlength}{2368sp}%
\begingroup\makeatletter\ifx\SetFigFont\undefined%
\gdef\SetFigFont#1#2#3#4#5{%
  \reset@font\fontsize{#1}{#2pt}%
  \fontfamily{#3}\fontseries{#4}\fontshape{#5}%
  \selectfont}%
\fi\endgroup%
\begin{picture}(6537,3300)(751,-3874)
\put(751,-886){\makebox(0,0)[lb]{\smash{{\SetFigFont{8}{9.6}{\rmdefault}{\mddefault}{\updefault}$X_n^{(1)}$}}}}
\put(751,-2161){\makebox(0,0)[lb]{\smash{{\SetFigFont{8}{9.6}{\rmdefault}{\mddefault}{\updefault}$X_n^{(K)}$}}}}
\put(2101,-3811){\makebox(0,0)[lb]{\smash{{\SetFigFont{8}{9.6}{\rmdefault}{\mddefault}{\updefault}$Y_n^{(1)}$}}}}
\put(3451,-3811){\makebox(0,0)[lb]{\smash{{\SetFigFont{8}{9.6}{\rmdefault}{\mddefault}{\updefault}$Y_n^{(K)}$}}}}
\put(2776,-2011){\makebox(0,0)[lb]{\smash{{\SetFigFont{10}{12.0}{\rmdefault}{\mddefault}{\updefault}and}}}}
\put(5551,-2086){\makebox(0,0)[lb]{\smash{{\SetFigFont{10}{12.0}{\rmdefault}{\mddefault}{\updefault}($K$-dim
VQ)}}}}
\put(2176,-1186){\makebox(0,0)[lb]{\smash{{\SetFigFont{10}{12.0}{\rmdefault}{\mddefault}{\updefault}$K$
independent}}}}
\put(6001,-1111){\makebox(0,0)[lb]{\smash{{\SetFigFont{10}{12.0}{\rmdefault}{\mddefault}{\updefault}joint}}}}
\put(5551,-1411){\makebox(0,0)[lb]{\smash{{\SetFigFont{10}{12.0}{\rmdefault}{\mddefault}{\updefault}quantization}}}}
\put(2176,-1561){\makebox(0,0)[lb]{\smash{{\SetFigFont{10}{12.0}{\rmdefault}{\mddefault}{\updefault}pre-post
filters}}}}
\put(2101,-2461){\makebox(0,0)[lb]{\smash{{\SetFigFont{10}{12.0}{\rmdefault}{\mddefault}{\updefault}prediction
loops}}}}
\put(4426,-811){\makebox(0,0)[lb]{\smash{{\SetFigFont{8}{9.6}{\rmdefault}{\mddefault}{\updefault}$\bZ$}}}}
\put(4501,-1711){\makebox(0,0)[lb]{\smash{{\SetFigFont{8}{9.6}{\rmdefault}{\mddefault}{\updefault}$\bZ
q$}}}}
\end{picture}%
\caption{DPCM of parallel sources.} \label{parallel_sources}
\end{figure}

One way we can achieve the VQ gain and still retain the
sequential structure of the system is by adding a
``spatial'' dimension, i.e., by jointly encoding a large number of
{\em parallel sources}, as happens, e.g., in video coding.
\figref{parallel_sources} shows DPCM encoding of $K$ parallel
sources. The spectral shaping and prediction are done in the time
domain for each source separately. Then, the
resulting vector of $K$ prediction errors is quantized {\em jointly}
at each time instant by a vector quantizer. The desired properties
of additive quantization error, and rate which is equal to $K$
times the mutual information $I(Z_n; Z_n + N_n)$, can be
approached in the limit of large $K$ by a suitable choice of the
quantizer.
In the next section we discuss
one way to do that using lattice ECDQ.

What if we have only one source instead of $K$ parallel sources?
If the source has decaying memory, we
can still approximate the parallel source coding approach
above, at the cost of large delay, by using {\em interleaving}.
We divide the (pre-filtered) source into $K$ long blocks,
which are separately predicted and then interleaved
and jointly quantized as if they were parallel sources.
See Figure~\ref{interleaved_DPCM}.
This is analogous to the method used in \cite{GuessVaranasiIT}
for combining coding-decoding and decision-feedback equalization
(DFE).

\begin{figure}[h]
\centering
\begin{picture}(0,0)%
\includegraphics{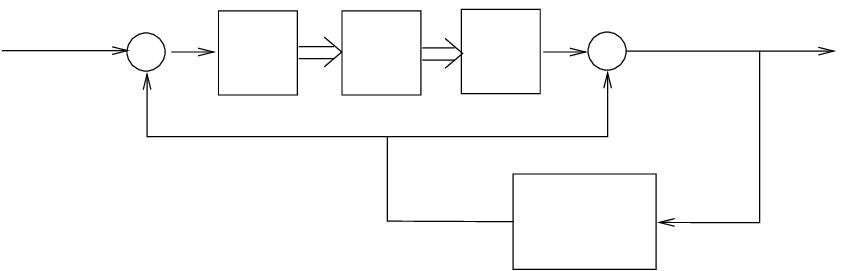}%
\end{picture}%
\setlength{\unitlength}{2408sp}%
\begingroup\makeatletter\ifx\SetFigFont\undefined%
\gdef\SetFigFont#1#2#3#4#5{%
  \reset@font\fontsize{#1}{#2pt}%
  \fontfamily{#3}\fontseries{#4}\fontshape{#5}%
  \selectfont}%
\fi\endgroup%
\begin{picture}(6634,2084)(1962,-1873)
\put(6226,-1561){\makebox(0,0)[lb]{\smash{{\SetFigFont{7}{8.4}{\rmdefault}{\mddefault}{\updefault}Predictor}}}}
\put(4801,-211){\makebox(0,0)[lb]{\smash{{\SetFigFont{7}{8.4}{\rmdefault}{\mddefault}{\updefault}VQ}}}}
\put(3012,-211){\makebox(0,0)[lb]{\smash{{\SetFigFont{8}{9.6}{\rmdefault}{\mddefault}{\updefault}$\Sigma$}}}}
\put(2810,-532){\makebox(0,0)[lb]{\smash{{\SetFigFont{7}{8.4}{\rmdefault}{\mddefault}{\updefault}$-$}}}}
\put(6515,-519){\makebox(0,0)[lb]{\smash{{\SetFigFont{7}{8.4}{\rmdefault}{\mddefault}{\updefault}$+$}}}}
\put(6665,-219){\makebox(0,0)[lb]{\smash{{\SetFigFont{8}{9.6}{\rmdefault}{\mddefault}{\updefault}$\Sigma$}}}}
\put(3840,-200){\makebox(0,0)[lb]{\smash{{\SetFigFont{10}{12.0}{\rmdefault}{\mddefault}{\updefault}$\Pi$}}}}
\put(5689,-221){\makebox(0,0)[lb]{\smash{{\SetFigFont{10}{12.0}{\rmdefault}{\mddefault}{\updefault}$\Pi^{-1}$}}}}
\put(1962,
67){\makebox(0,0)[lb]{\smash{{\SetFigFont{7}{8.4}{\rmdefault}{\mddefault}{\updefault}from
pre-filter}}}} \put(7627,
29){\makebox(0,0)[lb]{\smash{{\SetFigFont{7}{8.4}{\rmdefault}{\mddefault}{\updefault}to
post-filter}}}}
\end{picture}%
\caption{VQ-DPCM for a single source using interleaving. ($\Pi$ and
$\Pi^{-1}$ denote interleaving and de-interleaving, respectively.)}
\label{Interleaved_DPCM}
\end{figure}

If we do not use any of the above, but restrict ourselves to
scalar quantization ($K=1$), then we have a pre/post filtered DPCM scheme.
By combining Theorem~\ref{thm1} with known bounds on the performance of
(memoryless) entropy-constrained scalar quantizers
(e.g., \cite{FederZamir94}),
we have
\begin{equation}
\label{bound4scalar}
H(Q^{opt}(Z_n)) \leq R(D) + \frac{1}{2}
\log \Bigl( \frac{2 \pi e}{12} \Bigr)
\end{equation}
where
$1/2 \log(2 \pi e / 12) \approx 0.254$ bit.
See Remark~3 in the next section regarding scalar/lattice ECDQ.
Hence, Theorem~\ref{thm1} implies
that in principle, a pre/post
filtered DPCM scheme is optimal, up to the loss of the VQ gain,
at all distortion levels and not only at the high resolution
regime.

%
A different approach to combine VQ and prediction is first to
extract the innovation process and then to quantize it. It is
interesting to mention that this method of ``open loop'' prediction,
which we mentioned earlier regarding the model of
\cite{BergerD*PCM}, is known in the quantization literature as
D$^*$PCM  \cite{Jayant84}.
The best pre-filter for D$^*$PCM under a
high resolution assumption turns out to be the ``half-whitening
filter'': $|H_1(e^{j 2 \pi f})|^2 = 1/\sqrt{S(e^{j 2 \pi f})}$, with
the post filter $H_2(e^{j 2 \pi f})$ being its inverse.  But even
with this optimum filter, D$^*$PCM is inferior to DPCM: The optimal
distortion gain of D$^*$PCM over a non-predictive scheme is
\[
G_{\textrm{D$^*$PCM}} = \frac{\sigma_X^2}{\left( \int_{-1/2}^{1/2}\sqrt{S_X(e^{j 2 \pi
f})} df\right)^2}
\]
(strictly greater than one for non-white
spectra by the Cauchy-Schwartz inequality). Comparing to the optimum
prediction gain obtained by the DPCM scheme:
\[
G_{\textrm{DPCM}} =
\frac{\sigma_X^2}{P_e(X)} \ \ ,
\]
we have:
\[
\frac{G_{\textrm{DPCM}}}{G_{\textrm{D$^*$PCM}}} = \frac{\left( \int_{-1/2}^{1/2}\sqrt{S_X(e^{j 2 \pi
f})} df\right)^2}{P_e(X)} = \left(\frac{\sigma_{\tilde{U}}^2}{P_e(\tilde{U})}\right)^2
\ \ ,
\]
where $\tilde{U}_n$ is the pre-filter output in the D$^*$PCM scheme. This
ratio is strictly greater than one for non-white spectra.



\vspace{1cm}

\section{ECDQ in a Closed Loop System}
\label{ECDQ_sec}

\begin{figure}[!t]
\centering
\begin{picture}(0,0)%
\includegraphics{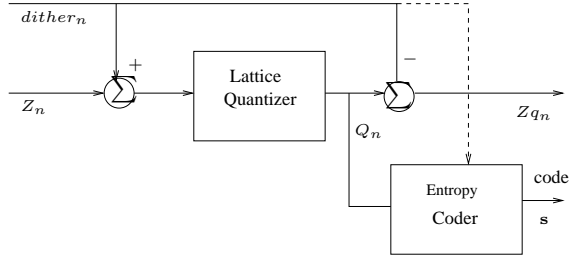}%
\end{picture}%
\setlength{\unitlength}{1973sp}%
\begingroup\makeatletter\ifx\SetFigFont\undefined%
\gdef\SetFigFont#1#2#3#4#5{%
  \reset@font\fontsize{#1}{#2pt}%
  \fontfamily{#3}\fontseries{#4}\fontshape{#5}%
  \selectfont}%
\fi\endgroup%
\begin{picture}(7042,3174)(2914,-2773)
\put(4426,-436){\makebox(0,0)[lb]{\smash{{\SetFigFont{6}{7.2}{\rmdefault}{\mddefault}{\updefault}$+$}}}}
\put(7876,-361){\makebox(0,0)[lb]{\smash{{\SetFigFont{6}{7.2}{\rmdefault}{\mddefault}{\updefault}$-$}}}}
\put(3076,164){\makebox(0,0)[lb]{\smash{{\SetFigFont{6}{7.2}{\rmdefault}{\mddefault}{\updefault}$dither_n$}}}}
\put(3076,-961){\makebox(0,0)[lb]{\smash{{\SetFigFont{6}{7.2}{\rmdefault}{\mddefault}{\updefault}$Z_n$}}}}
\put(5626,-886){\makebox(0,0)[lb]{\smash{{\SetFigFont{7}{8.4}{\rmdefault}{\mddefault}{\updefault}Quantizer}}}}
\put(5701,-586){\makebox(0,0)[lb]{\smash{{\SetFigFont{7}{8.4}{\rmdefault}{\mddefault}{\updefault}Lattice}}}}
\put(4201,-736){\makebox(0,0)[lb]{\smash{{\SetFigFont{6}{7.2}{\rmdefault}{\mddefault}{\updefault}$\sum$}}}}
\put(7651,-811){\makebox(0,0)[lb]{\smash{{\SetFigFont{6}{7.2}{\rmdefault}{\mddefault}{\updefault}$\sum$}}}}
\put(8251,-2386){\makebox(0,0)[lb]{\smash{{\SetFigFont{7}{8.4}{\rmdefault}{\mddefault}{\updefault}Coder}}}}
\put(8176,-2011){\makebox(0,0)[lb]{\smash{{\SetFigFont{6}{7.2}{\rmdefault}{\mddefault}{\updefault}Entropy}}}}
\put(9601,-2386){\makebox(0,0)[lb]{\smash{{\SetFigFont{6}{7.2}{\rmdefault}{\mddefault}{\updefault}$\bs$}}}}
\put(9301,-1036){\makebox(0,0)[lb]{\smash{{\SetFigFont{6}{7.2}{\rmdefault}{\mddefault}{\updefault}$Zq_n$}}}}
\put(7276,-1261){\makebox(0,0)[lb]{\smash{{\SetFigFont{6}{7.2}{\rmdefault}{\mddefault}{\updefault}$Q_n$}}}}
\put(9526,-1861){\makebox(0,0)[lb]{\smash{{\SetFigFont{7}{8.4}{\rmdefault}{\mddefault}{\updefault}code}}}}
\end{picture}%
\caption{ECDQ Structure.} \label{ECDQ_fig}
\end{figure}

Subtractive dithering of a uniform/lattice quantizer is a common
approach to make the quantization noise additive. As shown in
\cite{FederZamir}, the conditional entropy of the dithered
lattice quantizer (given the dither) is equal to the mutual information in
an additive noise channel, where the noise is uniform over the
lattice cell.   Furthermore, for ``good'' high dimensional
lattices, the noise becomes closer to a white Gaussian process \cite{LQN}.
Thus, ECDQ (entropy-coded dithered quantization)
provides a natural way to realize the inner AWGN channel block
of the predictive test-channel.

One difficulty, however, we observe in this section is that the results
developed in \cite{FederZamir} do not apply to the case where
the ECDQ input depends on previous EDCQ outputs and the entropy
coding is conditioned on the past.
This situation indeed happens in predictive coding, when ECDQ is
embedded within a feedback loop.
As we shall see, the right measure in this case
is the {\em directed information}.

An ECDQ operating on the source $Z_n$ is depicted in \figref{ECDQ_fig}.
A dither sequence $D_n$, independent of the input
sequence $Z_n$, is added before the quantization and subtracted
after.
If the quantizer has a lattice structure of dimension $K\geq 1$,
then we assume that the sequence length is
$$L = M K$$
for some integer $M$, so the quantizer is activated $M$ times. In
this section, we use bold notation for $K$-blocks corresponding to a
single quantizer operation. At each quantizer operation instant $m$,
a dither vector $\bD_m$ is independently and uniformly distributed
over the basic lattice cell. The lattice points at the quantizer
output ${\bf Q}_m, \ m = 1,\ldots,M$ are fed into an entropy coder
which is allowed to jointly encode the sequence, and has knowledge
of the dither as well, thus for an input sequence of length $L$ it
achieves an average rate of: \beq{R_ECDQ} R_{ECDQ} \Ddef \frac{1}{L}
H(\bQ_1^{M} | \bD_1^{M}) \ \ \eeq bit per source sample. The entropy
coder produces a sequence $\bs$ of $\left\lceil L
R_{ECDQ}\right\rceil$ bits, from which the decoder can recover
$\bQ_1,\ldots \bQ_M$, and then subtract the dither to obtain the
reconstruction sequence $Zq_n = Q_n - D_n, \ n = 1, \ldots L$. The
reconstruction error sequence
\[N_n = Zq_n - Z_n , \]
called in the sequel the ``ECDQ noise'',
has $K$-blocks which
are uniformly distributed over the mirror image of the basic lattice
cell and are mutually i.i.d. \cite{FederZamir}.
It is further stated in \cite[Thm.1]{FederZamir}
that the input and the noise sequences,
$\bZ = Z_1^L$ and $\bN = N_1^L$, are
statistically independent,
and that the ECDQ rate is equal to the
mutual information over an additive noise channel with the input $\bZ$ and
the noise $\bN$:
\beqn{R_no_feedback}
R_{ECDQ} & = & \frac{1}{L}
I(\bZ ; {\bf Zq} ) \nonumber \\
& = & \frac{1}{L} I (\bZ ; \bZ +  \bN ) \ \ \ . \eeqn

However,
the derivation of \cite[Thm. 1]{FederZamir} makes the implicit
assumption that the quantizer is \emph{used without feedback}, that
is, the current input is conditionally independent of past outputs
given the past inputs.  (In other words, the dependence
on the past, if exists, is only due to memory in the source.)
When there is feedback, this condition does
not necessarily hold, which implies that (even with the dither) the sequences $\bZ$
and $\bN$ are possibly dependent.
Specifically, since feedback is causal, the input $Z_n$ can depend
on past values of the ECDQ noise $N_n$, so their joint distribution
in general has the form:
\beqn{joint_feedback}
f( Z_1^L, N_1^L) =
\prod_{m=1}^{M} f( \bN_{m}) f(\bZ_{m} | \bN_{1}^{m-1}) \ \
\eeqn
where $$\bZ_m = Z_{(m-1)K+1}^{mK}$$ denotes
the $m$th $K$-block, and similarly for ${\bf N}_m$.
In this case, the
mutual information rate of \eqref{R_no_feedback} over-estimates the
true rate of the ECDQ.

Massey shows in \cite{Massey1990} that for DMCs with feedback,
traditional mutual information is not a suitable measure, and should
be replaced by \emph{directed information}. The directed information
between the sequences $\bZ$ and ${\bf Zq} = Zq_1^L$ is defined as
\beqn{directed}
I( \bZ \rightarrow {\bf Zq}) &\Ddef& \sum_{n=1}^{L} I
(Z_1^n ; Zq_n | Zq_1^{n-1}) \\
&=&
\sum_{n=1}^{L} I(Z_n ; Zq_n | Zq_1^{n-1}) \nonumber \ \
\eeqn
where the second equality holds whenever the channel from
$Z_n$ to $Zq_n$ is memoryless, as in our case.
In contrast, the mutual
information between $\bZ$ and ${\bf Zq}$ is given by,
\beq{directed2} I( \bZ; {\bf Zq}) = \sum_{n=1}^{L} I (Z_1^L ; Zq_n |
Zq_1^{n-1}) \eeq which by the chain rule for mutual information is
in general higher. For our purposes, we will define the $K$-block
directed information:
\beq{block_directed} I_K(\bZ \rightarrow {\bf Zq}) \Ddef
\sum_{m=1}^{M}  I( \bZ_1^{m} ;  {\bf Zq}_{m} | \bZ q_1^{m-1}) \ \
\eeq
%
The following result, proven in Appendix A,
extends Massey's observation to ECDQ with feedback,
and generalizes the result of \cite[Thm. 1]{FederZamir}:

\vspace{.3in}

\begin{theorem}
\label{thmECDQ}
{\bf (ECDQ Rate with Feedback)}
The ECDQ system with causal feedback defined by
\eqref{joint_feedback} satisfies: \beq {R_feedback} R_{ECDQ} =
\frac{1}{L} I_K(\bZ \rightarrow {\bf Zq})
= \frac{1}{L} I_K(\bZ \rightarrow \bZ + \bN )
 \ \ .
 \eeq
\end{theorem}

\vspace{.3in}

\textbf{Remarks:}

1. When there is no feedback, the past and future
input blocks $(\bZ_1^{m-1}, \bZ_{m+1}^M)$
are conditionally independent of the current output block
$\bZ q_m$ given the current input block $\bZ_m$,
implying by the chain rule that (\ref{directed}) coincides
with (\ref{directed2}),
and Theorem~\ref{thmECDQ} reduces to \cite[Thm. 1]{FederZamir}.

2. Even for scalar quantization ($K=1)$,
the ECDQ rate \eqref{R_ECDQ} refers to joint entropy coding of
the whole input vector.
This does not contradict the
sequential nature of the system since
entropy coding can be implemented causally.
Indeed, it follows from the chain rule for entropy that it is
enough to encode the instantaneous quantizer output
$\bQ_{m}$ conditioned on past quantizer outputs $\bQ_1^{m-1}$
and on past and present dither samples $\bD_1^{m}$,
in order to achieve the joint entropy of the quantizer
in (\ref{R_ECDQ}).

3. If we don't condition the entropy coding on the past,
then we have
\begin{eqnarray}
R_{ECDQ} &=& I(Z_n; Z_n + N_n^{(uniform)}) \\
&\leq&   I(Z_n; Z_n + N_n^{(gauss)})
+ \frac{1}{2} \log \Bigl( \frac{2 \pi e}{12} \Bigr)
\\
\label{xxx}
&=& R(D) + \frac{1}{2} \log \Bigl( \frac{2 \pi e}{12} \Bigr)
\end{eqnarray}
where
$N_n^{(uniform)}$, the scalar quantization noise,
is uniformly distributed over the interval
$(-\sqrt{12D}, +\sqrt{12D})$,
and where (\ref{xxx}) follows from Theorem~\ref{thm1}.
This implies (\ref{bound4scalar}) in the previous section.

4.  We can embed a $K$-dimensional lattice ECDQ for $K>1$
in the predictive test channel of
\figref{scheme_fig}, instead of the additive noise channel,
using the Vector-DPCM (VDPCM) configuration discussed in the
previous section.
For good lattices, when the quantizer dimension
$K \rightarrow \infty$, the noise process $\bN$
in the rate expressions
(\ref{R_no_feedback}) and (\ref{R_feedback})
becomes white Gaussian, and the scheme achieves the rate-distortion function.
Indeed,
combining Theorems~\ref{thm1} and~\ref{thmECDQ}, we see that
the average rate per sample of such VDPCM with ECDQ
satisfies:
\[ R_{VDPCM-ECDQ} = I(Z_n; Z_n + N_n) = R(D) \ \ .\]
This implies, in particular, that
the entropy coder does not need to be conditioned on the past
at all, as the predictor handles all the memory.
However, when the
quantization noise is not Gaussian, or the predictor is not optimal,
the entropy coder can use the residual time-dependence after
prediction to further reduce the coding rate.
The resulting rate of the ECDQ would be the average
directed information between the source and its reconstruction
as stated in Theorem~\ref{thmECDQ}.



\vspace{1cm}

\section{A Dual Relationship with Decision-Feedback Equalization}
\label{channel_sec}

In this section we make an analogy between the predictive form of the
Gaussian RDF and the ``information-optimality'' of decision-feedback
equalization (DFE) for colored Gaussian channels.
As we shall see,
a symmetric equivalent form of this channel coding problem,
including a water-pouring transmission filter, an MMSE
receive filter and a noise prediction feedback loop,
exhibits a striking resemblance to the pre/post-filtered
predictive test-channel of Figure~\ref{scheme_fig}.

\begin{figure*}[!t]
\centering
  \begin{psfrags}
  \psfrag{H}{"$+E_n$"}
     \psfrag{A}{$X_n$}
\psfrag{b1}{$\Sigma$} \psfrag{b2}{$\Sigma$} \psfrag{T}{$N_n$}
\psfrag{Y}{$Y_n$} \psfrag{b3}{$\Sigma$} \psfrag{b4}{$V_n$}
\psfrag{B}{$N_n$} \psfrag{C}{$\hat{U}_n$} \psfrag{D}{$\hat{U}_n$}
\psfrag{E}{$D_n$} \psfrag{F}{$\hat{D}_n$} \psfrag{G}{$U_n$}
     \includegraphics[scale=0.72]{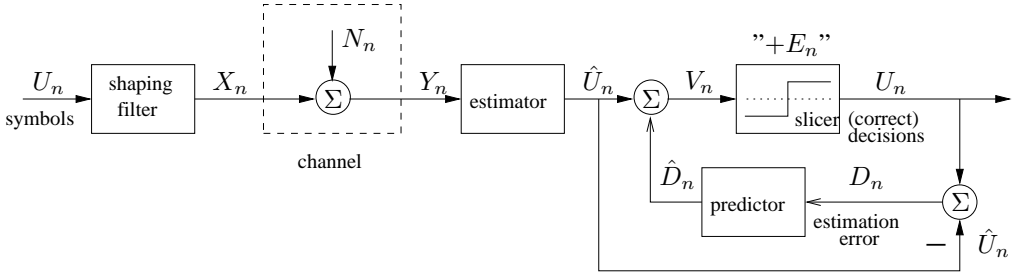}
  \end{psfrags}
 \caption{MMSE-DFE Scheme in Predictive Form}
   \label{dfefig}
\end{figure*}
We consider the (real-valued) discrete-time time-invariant linear
Gaussian channel,
\begin{equation}
R_n = c_n * X_n + Z_n, \label{channelISI}
\end{equation}
where the transmitted signal $S_n$ is subject to a power
constraint $E[S^2_n] \leq P$,
the channel dispersion is modeled by a linear time-invariant filter $c_n$,
and where the channel noise $Z_n$ is (possibly
colored) Gaussian noise.

Let $U_n$ represent the data stream which we model as an i.i.d.
zero-mean Gaussian random process with variance $\sigma_U^2$.
Further, let $h_{1,n}$ be a spectral shaping filter, satisfying
\begin{eqnarray}
\label{power_constraint}
\sigma_U^2 \int_{-1/2}^{1/2}|H_1(e^{j 2 \pi f})|^2 df  
\leq P
\end{eqnarray}
so the channel input $X_n = h_{1,n} * U_n$ indeed satisfies the
power constraint. For the moment, we make no further assumption on
$h_n$.

The channel (\ref{channel}) has inter-symbol interference (ISI) due
to the channel filter $c_n$, as well as colored Gaussian noise. Let
us assume that the channel frequency response is non-zero
everywhere, and pass the received signal $R_n$ through a
zero-forcing (ZF) linear equalizer $\frac{1}{C(z)}$,
resulting in $Y_n$. 
We thus arrive at an equivalent ISI-free channel,
\begin{equation}
Y_n=X_n+N_n, \label{channel}
\end{equation}
where the power spectrum of $N_n$ is
\[S_{N}(e^{j 2 \pi f})=\frac{S_{Z}(e^{j 2 \pi f})}{|C(e^{j 2 \pi f})|^2}.
\]


The mutual information rate (normalized per symbol) (\ref{mutual_info_rate})
between the input and output of the channel (\ref{channel}) is
\begin{eqnarray}
\oI(\{X_n\},\{Y_n\}) = \int_{-1/2}^{1/2} \frac{1}{2} \log \left(1+
\frac{ S_X(\ej)} {S_{N}(e^{j2\pi f})}\right)df. \label{MI}
\label{l9}
\end{eqnarray}
We note that if the spectral shaping filter $h_n$ satisfies the
optimum ``water-filling'' power allocation condition,
\cite{CoverBook},
then (\ref{l9}) will equal the channel capacity.

\begin{figure*}
\centering
\begin{picture}(0,0)%
\includegraphics{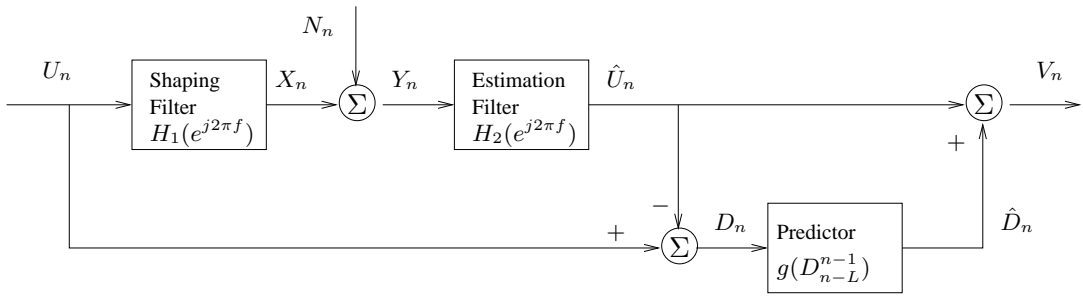}%
\end{picture}%
\setlength{\unitlength}{2960sp}%
\begingroup\makeatletter\ifx\SetFigFont\undefined%
\gdef\SetFigFont#1#2#3#4#5{%
  \reset@font\fontsize{#1}{#2pt}%
  \fontfamily{#3}\fontseries{#4}\fontshape{#5}%
  \selectfont}%
\fi\endgroup%
\begin{picture}(9144,2424)(3514,-1723)
\put(6376,-211){\makebox(0,0)[lb]{\smash{{\SetFigFont{11}{13.2}{\rmdefault}{\mddefault}{\updefault}$\Sigma$}}}}
\put(9076,-1411){\makebox(0,0)[lb]{\smash{{\SetFigFont{11}{13.2}{\rmdefault}{\mddefault}{\updefault}$\Sigma$}}}}
\put(11626,-211){\makebox(0,0)[lb]{\smash{{\SetFigFont{11}{13.2}{\rmdefault}{\mddefault}{\updefault}$\Sigma$}}}}
\put(3826,
89){\makebox(0,0)[lb]{\smash{{\SetFigFont{9}{10.8}{\rmdefault}{\mddefault}{\updefault}$U_n$}}}}
\put(5776,
14){\makebox(0,0)[lb]{\smash{{\SetFigFont{9}{10.8}{\rmdefault}{\mddefault}{\updefault}$X_n$}}}}
\put(8551,
14){\makebox(0,0)[lb]{\smash{{\SetFigFont{9}{10.8}{\rmdefault}{\mddefault}{\updefault}$\hat
U_n$}}}}
\put(6001,464){\makebox(0,0)[lb]{\smash{{\SetFigFont{9}{10.8}{\rmdefault}{\mddefault}{\updefault}$N_n$}}}}
\put(6751,
14){\makebox(0,0)[lb]{\smash{{\SetFigFont{9}{10.8}{\rmdefault}{\mddefault}{\updefault}$Y_n$}}}}
\put(9451,-1186){\makebox(0,0)[lb]{\smash{{\SetFigFont{9}{10.8}{\rmdefault}{\mddefault}{\updefault}$D_n$}}}}
\put(12151,
89){\makebox(0,0)[lb]{\smash{{\SetFigFont{9}{10.8}{\rmdefault}{\mddefault}{\updefault}$V_n$}}}}
\put(8926,-1036){\makebox(0,0)[lb]{\smash{{\SetFigFont{9}{10.8}{\rmdefault}{\mddefault}{\updefault}$-$}}}}
\put(11851,-1186){\makebox(0,0)[lb]{\smash{{\SetFigFont{9}{10.8}{\rmdefault}{\mddefault}{\updefault}$\hat
D_n$}}}}
\put(4726,-436){\makebox(0,0)[lb]{\smash{{\SetFigFont{9}{10.8}{\rmdefault}{\mddefault}{\updefault}$H_1(\ej)$}}}}
\put(7426,-436){\makebox(0,0)[lb]{\smash{{\SetFigFont{9}{10.8}{\rmdefault}{\mddefault}{\updefault}$H_2(\ej)$}}}}
\put(9976,-1561){\makebox(0,0)[lb]{\smash{{\SetFigFont{9}{10.8}{\rmdefault}{\mddefault}{\updefault}$g(D_{n-L}^{n-1})$}}}}
\put(8551,-1261){\makebox(0,0)[lb]{\smash{{\SetFigFont{9}{10.8}{\rmdefault}{\mddefault}{\updefault}$+$}}}}
\put(11401,-511){\makebox(0,0)[lb]{\smash{{\SetFigFont{9}{10.8}{\rmdefault}{\mddefault}{\updefault}$+$}}}}
\end{picture}%
\caption{Noise-Prediction Equivalent Channel} \label{channelfig}
\end{figure*}

%

Similarly to the observations made in Section~\ref{intro} with
respect to the RDF, we note (as reflected in \eqref{l9}) that
capacity may be achieved by parallel AWGN coding over narrow
frequency bands (as done in practice in Discrete Multitone
(DMT)/Orthogonal Frequency-Division Multiplexing (OFDM) systems). An
alternative approach, based on time-domain prediction rather than
the Fourier transform, is offered by the canonical MMSE - feed
forward equalizer - decision feedback equalizer (FFE-DFE) structure
used in single-carrier transmission. It is well known that this
scheme, coupled with AWGN coding, can achieve the capacity of linear
Gaussian channels. This has been shown using different approaches by
numerous authors; see \cite{GuessVaranasiIT}, \cite{MMSE-DFE},
\cite{LeeMesserBOOK}, \cite{Forneyallerton04} and references
therein. Our exposition closely follows that of Forney
\cite{Forneyallerton04}. We now recount this result, based on linear
prediction of the error sequence; see the system in
Figure~\ref{dfefig} and its equivalent channel in
Figure~\ref{channelfig}. In the communication literature, this
structure is referred to as ``noise prediction". It can be recast
into the more familiar FFE-DFE form by absorbing a part of the
predictor into the estimator filter, forming the usual FFE.

As a first step, let $\hat{U}_n$ be the optimal MMSE estimator of
$U_n$ from the equivalent channel output sequence $\{Y_n\}$ of
(\ref{channel}). Since $\{U_n\}$ and $\{Y_n\}$ are jointly Gaussian
and stationary this estimator is linear and time invariant. Note
that the combination of the ZF equalizer $\frac{1}{C(z)}$ at the
receiver front-end and the estimator above is equivalent to direct
MMSE estimation of $U_n$ from the original channel output $R_n$
(\ref{channelISI}).
%
%
Denote the estimation error, which is composed in general of ISI and
Gaussian noise, by $D_n$. Then
\begin{equation}
U_n=\hat{U}_n+D_n
\end{equation}
where $\{D_n\}$ is statistically independent of $\{\hat{U}_n\}$ due to the
orthogonality principle and Gaussianity.


Assuming the decoder has access to past symbols
$U_n^- = U_{n-1},U_{n-2},\ldots$
(see in the sequel),
the decoder knows also the past estimation errors
$D_n^- = D_{n-1},D_{n-2},\ldots$ and may
form an optimal linear predictor, $\hat{D}_n$, of the current estimation
error $D_n$, which may then be added to $\hat U_n$ to form $V_n$.
The prediction error $E_n=D_n-\hat{D}_n$ has variance $P_e(D)$, the
entropy power of $D_n$.
It follows that
\begin{eqnarray}
U_n & = & \hat{U}_n+D_n \nonumber \\
         & = & V_n-\hat{D}_n+D_n \nonumber \\
         & = & V_n+E_n ,
         \label{backchannel}
\end{eqnarray}
and therefore
\begin{eqnarray}
\label{slicer_error}
E\{ U_n- V_n \}^2 = \sigma_E^2 = E\{ D_n- \hat{D}_n \}^2 =  P_e(D).
\end{eqnarray}

The channel (\ref{backchannel}), which describes the input/output
relation of the slicer in Figure~\ref{dfefig}, is often referred to
as the {\em backward channel}. Furthermore, since $U_n$ and $E_n$
are i.i.d Gaussian and since by the orthogonality principle $E_n$ is
independent of present and past values of $V_n$ (but dependent of
future values through the feedback loop), it is a ``sequentially
additive'' AWGN channel. See \figref{orthogonality_fig} for a
geometric view of these properties. Notice the strong resemblance
with the channel (\ref{coreAWGN}), $Zq_n = Z_n + N_n$, in the
predictive test-channel of the RDF: in both channels the output and
the noise are i.i.d. and Gaussian, but the input has memory and it
depends on past outputs via the feedback loop.

\begin{figure}[h]
\centering
\begin{picture}(0,0)%
\includegraphics{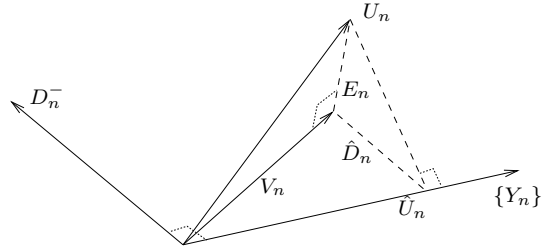}%
\end{picture}%
\setlength{\unitlength}{2763sp}%
\begingroup\makeatletter\ifx\SetFigFont\undefined%
\gdef\SetFigFont#1#2#3#4#5{%
  \reset@font\fontsize{#1}{#2pt}%
  \fontfamily{#3}\fontseries{#4}\fontshape{#5}%
  \selectfont}%
\fi\endgroup%
\begin{picture}(4594,2204)(5938,-2998)
\put(10286,-2601){\makebox(0,0)[lb]{\smash{{\SetFigFont{8}{9.6}{\rmdefault}{\mddefault}{\updefault}$\{Y_n\}$}}}}
\put(9431,-2718){\makebox(0,0)[lb]{\smash{{\SetFigFont{8}{9.6}{\rmdefault}{\mddefault}{\updefault}$\hat
U_n$}}}}
\put(8190,-2517){\makebox(0,0)[lb]{\smash{{\SetFigFont{8}{9.6}{\rmdefault}{\mddefault}{\updefault}$V_n$}}}}
\put(8919,-1646){\makebox(0,0)[lb]{\smash{{\SetFigFont{8}{9.6}{\rmdefault}{\mddefault}{\updefault}$E_n$}}}}
\put(9112,-950){\makebox(0,0)[lb]{\smash{{\SetFigFont{8}{9.6}{\rmdefault}{\mddefault}{\updefault}$U_n$}}}}
\put(6121,-1721){\makebox(0,0)[lb]{\smash{{\SetFigFont{8}{9.6}{\rmdefault}{\mddefault}{\updefault}$D_n^-$}}}}
\put(8919,-2223){\makebox(0,0)[lb]{\smash{{\SetFigFont{8}{9.6}{\rmdefault}{\mddefault}{\updefault}$\hat
D_n$}}}}
\end{picture}%
\caption{Geometric View of the Estimation Process}
\label{orthogonality_fig}
\end{figure}

We have therefore derived the following.

\vspace{.3in}

\begin{theorem}
\label{thmC}
{\bf (Information Optimality of Noise Prediction)}
For stationary Gaussian processes $U_n$ and $N_n$, and
if $H_2(\ej)$ is chosen to be the optimal estimation filter of $U_n$
from $Y_n$ and the predictor $g(\cdot)$ is chosen to be the optimal
prediction filter of $D_n$ (with $L\rightarrow\infty$), then the
mutual information-rate (\ref{MI}) of the channel from $X_n$ to
$Y_n$ (or from $U_n$ to $Y_n$)
is equal to the
\emph{scalar} mutual information
\[I(V_n;V_n+E_n)\]
of the channel (\ref{backchannel}). Furthermore, if $H_1(\ej)$ is
chosen such that $S_X(\ej)$ equals the water-filling spectrum of the
channel input, then this mutual information equals the channel capacity.
\end{theorem}

\vspace{.3in}

\begin{proof}
Let $U_n^-=\{U_{n-1}, U_{n-2}, \ldots\}$ and $D_n^-=\{D_{n-1},
D_{n-2}, \ldots\}$. Using the chain rule of mutual information we
have
\begin{eqnarray}
\oI(\{U_n\},\{Y_n\})
& = & \oh(\{U_n\})-\oh(\{U_n\}|\{Y_n\}) \nonumber \\
& = & \oh(\{U_n\})-h(U_n|\{Y_n\}, U^-_n) \nonumber \\
& = & \oh(\{U_n\})-h(U_n-\hat{U}_n|\{Y_n\}, U^-_n) \nonumber \\
& = & \oh(\{U_n\})-h(D_n|\{Y_n\}, U^-_n)  \nonumber \\
& = & \oh(\{U_n\})-h(D_n|\{Y_n\}, D^-_n) \nonumber \\
& = & \oh(\{U_n\})-h(D_n-\hat{D}_n|\{Y_n\}, D^-_n) \nonumber \\
& = & \oh(\{U_n\})-h(E_n|\{Y_n\}, D^-_n) \nonumber \\
& = & \oh(\{U_n\})-h(E_n) \label{l8} \\
& = & I(V_n;V_n+E_n), \nonumber \label{new}
\end{eqnarray}
where
$\oh(\cdot)$
denotes differential entropy rate,
and where
(\ref{l8}) follows from successive application of the
orthogonality principle \cite{Forneyallerton04}, since we assumed
optimum estimation and prediction filters, which are MMSE
estimators in the Gaussian setting.
\end{proof}

In view of (\ref{MI}) and (\ref{slicer_error}),
and since $\oI(\{U_n\},\{Y_n\}) = \oI(\{X_n\},\{Y_n\})$,
Theorem~\ref{thmC} can be re-written as
\begin{eqnarray}
\int_{-1/2}^{1/2} \frac{1}{2}\log \left(1+ \frac{ S_X(\ej) }
{S_{N}(e^{j 2 \pi f})}\right)df = \frac{1}{2}  \log \left(\frac{
\sigma_U^2 } {\sigma_E^2}\right)  \label{l10}
\end{eqnarray}
%
%
from which we obtain the following well known formula for the
``SNR at the slicer'' for infinite order FFE-DFE,
\cite{MMSE-DFE,LeeMesserBOOK},
\begin{eqnarray}
\frac{ \sigma_U^2 } {\sigma_E^2}
& = &
\exp \left( \int_{-1/2}^{1/2} \log
\left(
1+ \frac{ S_X(\ej) }{S_{N}(e^{j 2 \pi f})}
 \right)df \right). \nonumber
\end{eqnarray}
%
%

We make a few remarks and interpretations regarding
the capacity-achieving predictive configuration,
which further enhance its duality relationship with the predictive
realization of the RDF.


\textbf{Slicing and Coding}
We assumed that the decoder has access
to past symbols. In the simplest realization, this is achieved by a decision
element (``slicer'') that works on a symbol-by-symbol basis.
In practice however, to approach capacity, the slicer must be replaced by
a ``decoder". Here we must actually break with the assumption that $X_n$ is a
Gaussian process. We implicitly assume that $X_n$ are symbols of a capacity-
achieving AWGN code. The slicer should be viewed as a mnemonic aid where in
practice an optimal decoder should be used.

However, we encounter two problems with this interpretation.  First, the
common view of a slicer is as a nearest neighbor quantizer.  Thus in order
to function correctly, the noise $E_n$ in (\ref{backchannel}) must be independent of the
symbols $U_n$ and not of the estimator $V_n$ (i.e., the channel should
be "forward" additive: $V_n = U_n + E_n$). This can be achieved by dithering
the codebook via a modulo-shift as in \cite{ErezZamirAWGN}.
This is reminiscent to the dithered quantization approach of
Section~\ref{ECDQ_sec}.
Another difficulty is the conflict between the inherent
decoding delay of a good code, and the sequential nature
of the noise-prediction DFE configuration.
Again (as with vector-DPCM in \secref{DPCM_sec}),
this may in principle be solved by incorporating an interleaver as suggested by Guess
and Varanasi \cite{GuessVaranasiIT}.

\textbf{Capacity achieving shaping filter.}
For any spectral
shaping filter $h_{1,n}$, the mutual information is given by
\eqref{l9}. The shaping filter $h_n$ which maximizes the mutual
information (and yields capacity) under the power constraint
(\ref{power_constraint}) is given by the parametric
water-filling formula: \beq{water_filling_c} \sigma_U^2 |H_1(e^{j 2
\pi f})|^2 = [ \theta -  S_{N}(e^{j 2 \pi f}) ]^+,
 \eeq
where the ``water level'' $\theta$ is chosen so that the power
constraint is met with equality,
\begin{eqnarray}
 \sigma_X^2 & = & \int_{-1/2}^{1/2} \sigma_U^2 |H(e^{j 2 \pi f})|^2 df \nonumber \\
 & =  & \int_{-1/2}^{1/2} \lceil \theta - S_{N}(e^{j 2 \pi
f})\rceil^+df = P. \label{water_filling_c2}
 \end{eqnarray}
Using this choice, and arbitrarily setting
\beq{sigmaU} \sigma_U^2 = \theta
\eeq
it can be verified that the shaping filter $H_1(\ej)$
and the estimation filter $H_2(\ej)$ satisfy the same complex
conjugate relation as the RDF-achieving pre- and post-filters
\eqref{prefilter} and  \eqref{postfilter}
\[
H_2(\ej) = H_1^*(\ej) .
\]
Under the same choice, we also have that:
\beq{backward_spectrum}
S_D(\ej) = \min \left\{ S_N(\ej),\theta \right\} \
\ .
\eeq


\textbf{Shaping, estimation and prediction at high SNR.}
At high signal-to-noise ratio (SNR), the shaping filter $H_1$ and
the estimation filter $H_2$ become all-pass, and can be replaced by
scalar multipliers. If we set the symbol variance as in
(\ref{sigmaU}), then we get at high SNR $\sigma_U^2 \approx P$, so
$X_n \approx U_n$ and $\hat{U}_n \approx Y_n$.
It follows that the estimation error $D_n \approx N_n$,
and therefore the slicer error $E_n$ becomes simply the
prediction error (or the entropy power) of the channel noise $N_n$.
This is the well known ``zero-forcing DFE'' solution
for optimum detection at high SNR
\cite{LeeMesserBOOK}.
We shall next see that the same behavior of the slicer error
holds even for non-asymptotic conditions.


\textbf{The prediction process when the Shannon upper bound is tight.}
The Shannon upper bound (SUB) on capacity states that
\begin{eqnarray}
C & \leq &  \frac{1}{2} \log(2 \pi e \sigma_Y^2)- \oh(N) \nonumber \\
& \leq & \frac{1}{2} \log\left(\frac{P + \sigma_{N}^2}{P_e(N)}
 \right) \Ddef C_{\rm SUB}, \label{sup}
\end{eqnarray}
where
$$\sigma_{N}^2 =
\int_{-1/2}^{1/2}  S_{N}(e^{j 2 \pi f})  df$$ is the variance of the
equivalent noise, and where equality holds if and only if the output
$Y_n$ is white.
This in turn is satisfied if and only if
\[\theta \geq \max_f  S_{N}(\ebj), \]
in which case $\theta = P + \sigma_N^2$.

If we choose $\sigma_U^2$ according to \eqref{sigmaU}, we have:

\begin{itemize}
    \item{The shaping and estimation filters satisfy
    \[ | H_1(\ej) |^2 = |H_2(\ej)|^2 = 1 - \frac{ S_N(e^{j 2 \pi f}) }{\theta} .\]}
    \item{$U_n$ and $Y_n$ are white, with the same variance $\theta$.} 
    \item{$X_n$ and $\hat{U}_n$ have the same power spectrum, $\theta~-~S_N(e^{j 2 \pi f})$.}
    \item{The power spectrum of $D_n$ is equal to the power spectrum
    of the noise $N_n$, $S_N(\ej)$. Consequently, the variance of $E_n$ which is equal to the
    entropy-power of $D_n$, is equal to $P_e(N)$.}
    \item{As a consequence we have
    \begin{eqnarray*}
    I(V_n; V_n+E_n) &=& h(U_n) - h(E_n)  \\
    &=& h\Bigl( \cN(0,\theta) \Bigr) - h\Bigl( \cN(0,P_e(N)) \Bigr) \\
    &=&  \frac{1}{2}  \log\Bigl( \frac{P+\sigma_N^2}{P_e(N)} \Bigr)
    \end{eqnarray*}
    which is indeed the SUB (\ref{sup}).
    }
    \end{itemize}

\textbf{An alternative derivation of Theorem~\ref{thmC} in the
spectral domain.} Similarly to the alternative proof of
\thref{thm1}, one can prove \thref{thmC} using the spectra derived
above.


\vspace{1cm}

\section{Summary}
We demonstrated the dual role of prediction in rate-distortion theory
of Gaussian sources and capacity of ISI channels.
These observations shed light on
the configurations of DPCM (for source compression)
and FFE-DFE (for channel demodulation), and show that
in principle they are ``information lossless''
for any distortion / SNR level.
The theoretic bounds, RDF and capacity, can be approached in practice by
appropriate use of feedback and linear estimation in the time domain
combined with coding across the ``spatial'' domain.

A prediction-based system has in many cases a delay lower than
that of a frequency domain approach, as is well known in practice.
We slightly touched on this issue when discussing the 0.5 bit
loss due to avoiding the (``non-causal'') pre/post filters.
But the full potential of this aspect requires further study.

It is tempting to ask whether the predictive form of the RDF
can be extended to more general sources and distortion measures
(and similarly for capacity of more general ISI channels).
Yet, examination of the arguments in our derivation reveals that it
is strongly tied to the quadratic-Gaussian case:
\begin{itemize}
\item The orthogonality principle, implied by the MMSE criterion,
guarantees the whiteness of the noisy prediction error $Zq_n$ and
its orthogonality with the past.
\item Gaussianity implies that
orthogonality is equivalent to statistical independence.
\end{itemize}
For other error criteria and/or non-Gaussian sources, prediction
(either linear or non-linear) is in general unable to remove
the dependence on the past.  Hence the scalar mutual information
over the prediction error channel would in general be greater
than the mutual information rate of the source before prediction.


\vspace{1cm}

\section*{Acknowledgement}
We'd like to thank Robert M. Gray for pointing to us the origin of DPCM
in a U.S. patent by C.C. Cutler in 1952.

\vspace{1cm}




\section*{Appendix}


\section*{A. Proof of \thref{thmECDQ}}
\label{AppThm2}

It will be convenient to look at $K$-blocks, which we denote by bold
letters as in \secref{ECDQ_sec}. Substituting the ECDQ rate
definition \eqref{R_ECDQ} and the $K$-block directed information
definition \eqref{block_directed}, the required result
\eqref{R_feedback} becomes:
\[ H(\bQ_1^{M} | \bD_1^{M}) = \sum_{m=1}^{M} I ( \bZ^m ; \bZ q_m |
\bZ q_m^{m-1}) \ \ \ . \] Using the chain rule for entropies, it is
enough to show that: \[ H( \bQ_m |
 \bQ_1^{m-1},  \bD_1^{M-1}) = I ( \bZ q_m ; \bZ_1^m |  \bZ q_1^{m-1}) \ \ \ . \] To that end, we have the
following sequence of equalities: \beqn{beqn1} & & H( \bQ_m |
 \bQ_1^{m-1},  \bD_1^{M-1}) \nonumber \\
&\stackrel{(a)}=&
 H( \bQ_m |
\bQ_1^{m-1},  \bD_1^m) \nonumber
\\ &\stackrel{(b)}=&
H( \bQ_m |  \bQ_1^{m-1},  \bD_1^m) - H ( \bQ_m |  \bQ_1^{m-1},
\bZ_1^m,
 \bD_1^m ) \nonumber
\\ & = & I( \bQ_m ;  \bZ_1^m |  \bQ_1^{m-1},
 \bD_1^m) \nonumber \\ &\stackrel{(c)}=&
 I( \bQ_m -  \bD_m ;
\bZ_1^m |  \bQ_1^{m-1},  \bD_1^m) \nonumber
\\ & = & I( \bZ q_m ;  \bZ_1^m |  \bQ_1^{m-1},  \bD_1^m)
\nonumber \\ & = & I( \bZ q_m ;  \bZ_1^m |  \bQ_1^{m-1}-\bD_1^{m-1},
\bD_1^m) \nonumber \\ & = & I( \bZ q_m ;  \bZ_1^m | \bZ q_1^{m-1},
\bD_1^m) \nonumber \\ &\stackrel{(d)}=& I( \bZ q_m ;  \bZ_1^m | \bZ
q_1^{m-1}, \bD_m) \nonumber \\
&=& I(\bZ q_m, \bD_m; \bZ_1^m | \bZ q_1^{m-1}) - I(\bD_m;\bZ_1^m |
\bZ q_1^{m-1}) \nonumber \\ &\stackrel{(e)}=& I(\bZ q_m; \bZ_1^m
| \bZ q_1^{m-1}) - I(\bD_m;\bZ_1^m | \bZ q_1^{m-1}) \nonumber \\
&\stackrel{(f)}=& I( \bZ q_m;  \bZ_1^m |  \bZ q_1^{m-1}) \ \ .
\nonumber \eeqn In this sequence, equality (a) comes from the
independent dither generation and causality of feedback. (b) is
justified because $\bQ_m$ is a deterministic function of the
elements on which the subtracted entropy is conditioned, thus
entropy is $0$. In (c) we subtract from the left hand side argument
of the mutual information one of the variables upon which mutual
information is conditioned. (d) and (e) hold since each dither
vector $\bD_m$ is a deterministic function of the corresponding
quantizer output $\bZ q_m$. Finally, (f) is true since $\bZ_1^m$ is
independent of $\bD_m$ (both conditioned on past quantized values
and unconditioned).



\end{document}